\documentclass[galaxies,review,accept,moreauthors,pdftex]{mdpi}
\firstpage{1}
\pubvolume{7}
\issuenum{3}
\articlenumber{79}
\pubyear{2019}
\copyrightyear{2019}
\history{Received: 31 July 2019; Accepted: 13 September 2019; Published: 18 September 2019}
\updates{yes} 

\Title{The History Goes On: Century Long Study of  Romano's Star $^{\dagger}$}
\usepackage{upgreek}
\newcommand{\Msun}{\ensuremath{\rm M_\odot}}

\newcommand{\Rsun}{\ensuremath{\rm R_\odot}}
\newcommand{\magn}[1]{^{\rm m}\!\!\!#1\,}
\newcommand{\arcsec}{\hbox{$^{\prime\prime}$}}

\newcommand{\sevensize}{}
\usepackage{booktabs}

\Author{Olga Maryeva $^{1,2,}$*\orcidA, Roberto F. Viotti $^{3}$,  Gloria Koenigsberger $^{4}$\orcidB,        Massimo Calabresi $^{5}$, Corinne~Rossi $^{6}$ and Roberto Gualandi $^{7}$}

\AuthorNames{Olga Maryeva, Roberto F. Viotti, Gloria Koenigsberger, Massimo Calabresi, Corinne Rossi and Roberto Gualandi}

\address{$^{1}$ \quad Astronomical Institute of the Czech Academy of Sciences, Fri\v{c}ova 298, 25165 Ond\v{r}ejov, Czech Republic\\
$^{2}$ \quad Sternberg Astronomical Institute, Lomonosov Moscow State University, Universitetsky pr. 13, 119234~Moscow, Russia \\
$^{3}$ \quad INAF-Istituto di Astrofisica e Planetologia Spaziali di Roma (IAPS-INAF), Via del Fosso del Cavaliere 100, 00133 Roma, Italy; roberto.viotti@iaps.inaf.it \\
$^{4}$ \quad Instituto de Ciencias F\'{\i}sicas, Universidad Nacional Aut\'onoma de M\'exico, Ave. Universidad S/N,  62210~Cuernavaca, Mexico; gloria@astro.unam.mx  \\
$^{5}$ \quad Associazione Romana Astrofili, Via Carlo Emanuele~I, n$^{\circ}$12A, 00185  Roma, Italy; m.calabresi@mclink.it \\
$^{6}$ \quad Physics Department, Universit\`a di Roma ``La Sapienza'', Piazza le Aldo Moro 5, 00185 Roma, Italy; corinne.rossi@uniroma1.it\\ 
$^{7}$ \quad INAF-Osservatorio Astronomico di Bologna, Via Ranzani 1, I-40127 Bologna, Italy; roberto.gualandi@inaf.it}

\corres{Correspondence: olga.maryeva@gmail.com}

\firstnote{Based on observations made with the Gran Telescopio Canarias (GTC), installed at the Spanish Observatorio del Roque de los Muchachos of the Instituto de Astrofisica de Canarias, in the island of La Palma,  and with the Cassini 1.52-m telescope of the Bologna Observatory (Italy), as well as data retrieved from the public archive of the  Special Astrophysical observatory of Russian Academy of Sciences (SAO RAS).}%

\abstract{GR\,290 (M\,33~V0532 = Romano's Star) is a unique  variable star in the M33 galaxy, which simultaneously displays variability typical for luminous blue variable (LBV) stars and physical parameters typical for nitrogen-rich Wolf-Rayet (WR) stars (WN). As of now, GR\,290 is the first object which is confidently classified as a post-LBV star.
In this paper, we outline the main results achieved from extensive photometric and spectroscopic observations of the star: the structure and chemical composition of its wind and its evolution over time, the systematic increase of the bolometric luminosity during the light maxima, the circumstellar environment. These results show that the current state of Romano's Star constitutes a fundamental link in the evolutionary path of very massive~stars.}

\begin{document}

\section{Introduction}

GR\,290 (M\,33 V0532 = Romano's Star)\footnote{The object has coordinates {$\alpha=01:35:09.701$}, $\delta=+30:41:57.17$ at J2000 epoch.} is a variable star in M\,33 galaxy discovered by Giuliano \citet{romano} who originally constructed its light curve and classified it as a Hubble-Sandage variable based on its photometric properties.
Later, in 1984, Peter Conti \citep{Conti} introduced a new class of objects which assimilated Hubble-Sandage variables---luminous blue variables (LBV), and~thus GR\,290 became an LBV candidate~\cite{HumphreysDavidson, szeifert}. This classification has later been supported by the spectroscopic~\citep{fabrika} and photometric~\citep{Kurtev2001} studies, as~well as by its large bolometric luminosity~\citep{Polcaro2003}. However, some arguments suggest that the objects is rather on a post-LBV stage already~\citep{polcaro10, Humphreys2014, Polcaro2016}.

{ Romano's star displays both strong spectral and photometric variability, with~several significant (about 1.5--2~mag) increases of brightness detected during its long monitoring (Polcaro et al.~\cite{Polcaro2016} and references therein). Such  variability is typical for LBV stars, while in the Hertzsprung-Russell (H-R)  GR\,290 lies in Wolf-Rayet (WR) stars region,  beyond~LBV instability strip \citep{Polcaro2016}. GR\,290 is presently in a short, and~thus very rare, transition phase between the LBV evolutionary phase and the nitrogen rich WR stellar class (WN). }
%
It is an extremely important target for studies of massive star evolution, especially the evolutionary link between LBVs, WR stars and supernovae (SNe).

In this paper we summarise the main results achieved in the study of Romano's star. We combine new studies of GR\,290's vicinity (Section~\ref{vicinity}) with its updated century-long photometric light curve (Section~\ref{photometry}). Then, based on spectral data, numerical simulations of its stellar atmosphere (Section~\ref{spectroscopy}) and the nebula surrounding it (Section~\ref{nebula}), we discuss the current evolutionary stage of the star in Section~\ref{conclusion}.


\section{Stellar Vicinity of Romano's~Star}\label{vicinity}

\begin{figure}[H]
\centering
\includegraphics[width=0.7\linewidth,clip,trim=45 25 0 0]{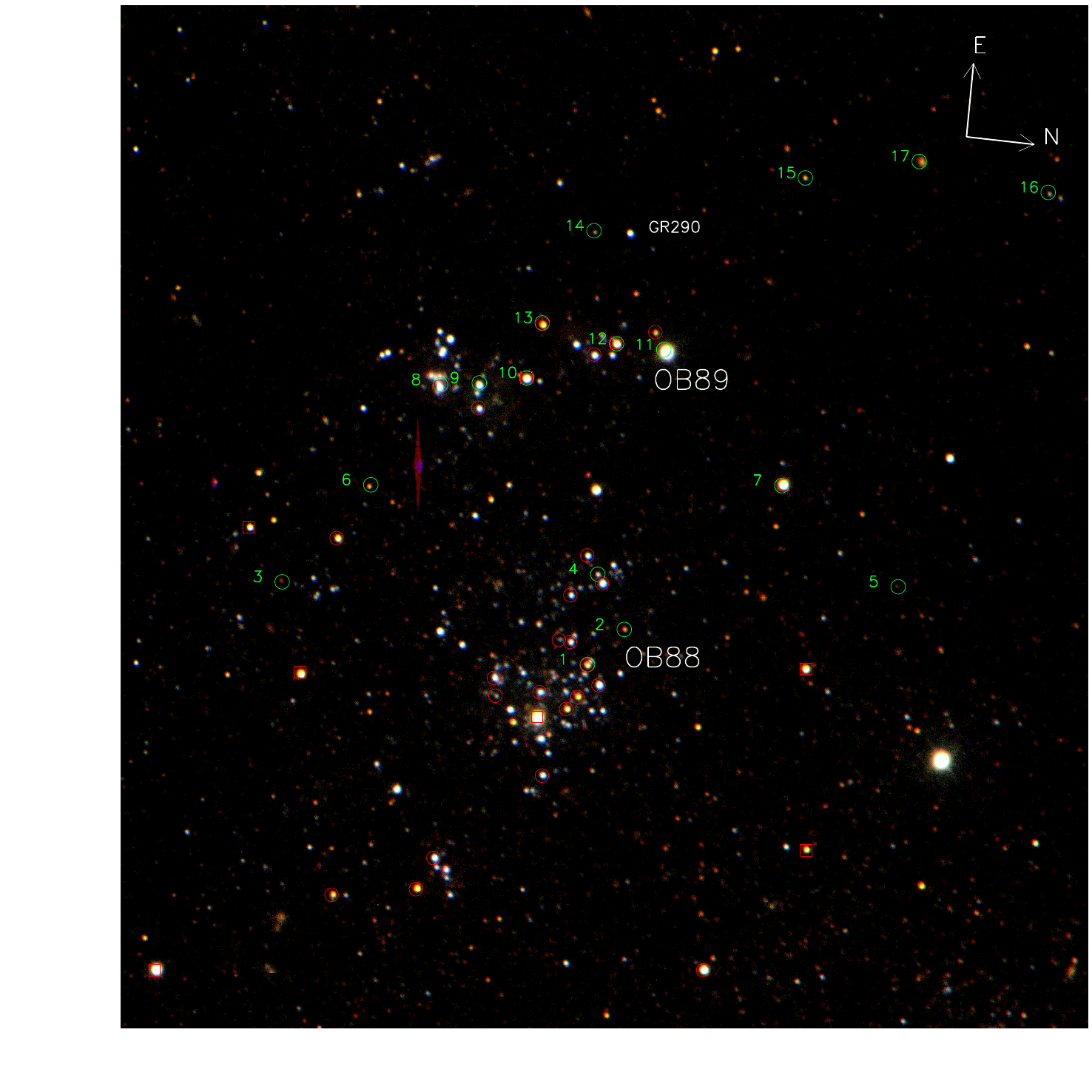}
\caption{Identification chart of GR\,290 vicinity and OB\,88 and OB\,89 associations. The~colour picture is a combination of three direct images, with~blue corresponding to B filter, green---to V and red---to R filter,
all obtained with 2.5\,m telescope of the Caucasian Mountain Observatory (CMO) of the Sternberg Astronomical Institute of Moscow State University. Green circles mark the stars studied in this work and red ones studied by Massey~et~al.~\cite{Massey2006,Massey2016ubv}. Red squares are stars considered to be foreground objects by Massey~et~al.~\cite{Massey2016ubv}.}
\label{fig_map}
\end{figure}

GR\,290 is located in the outer spiral arm of the M\,33 galaxy, and~lies to the east of the OB\,88 and OB\,89 associations  \citep{HumphreysSandage, Ivanov, MasseyArmandroff1995}, located at 0.5 and 0.125~kpc projected distances\footnote{The adopted distance to M\,33 is $847\pm61$~kpc (distance module $24.64\pm0.15$) from Galleti et al.~\cite{Galleti2004}.}, respectively. The~most detailed information about photometry of stars in this area may be found in Massey~et~al.~\cite{Massey2016ubv}. Figure~\ref{fig_map} shows the identification chart of the object and its vicinity, with~red symbols corresponding to the stars which were spectrally classified by Massey et al.~\cite{Massey2006}. Coordinates and spectral classes of the stars are listed in Table~\ref{tab:masseystars}.

\citet{MasseyJohnson1998}  found a couple of carbon-rich Wolf-Rayet (WC) stars in these associations, J013458.89+304129.0 (WC4) in OB\,88 and   J013505.37+304114.9 (WC4-5) in OB\,89. Moreover, the~OB\,88 association contains the star J013500.30+304150.9  classified as an LBV candidate by Massey~et~al.~\cite{Massey2007cLBV}, and~later reclassified by Humphreys et al.~\cite{Humphreys2014} as a FeII emission-line star. The~presence of evolved massive stars in the associations indicates that their age is close to that of GR\,290 and that they might have a common origin.
Therefore, it is quite reasonable to suppose that GR\,290 might have been originally ejected from the OB\,89 association. Then, assuming a median escape velocity for runaway stars of 40--200 km/s  \citep{Perets2012}, this ejection would have to have occurred 3.0--0.6
Myr ago, which~is consistent with the evolutionary age of GR\,290 and with the age of the OB\,89 association.
\begin{table}[H]
\centering
\caption{Stars in the vicinity of GR\,290 spectrally classified  by  Massey~et~al.~\cite{Massey2006} (and references therein).  Names and coordinates are given according to Massey~et~al.~\cite{Massey2016ubv}. The~three stars also included in Table~\ref{tab:nearstars}  are marked by boldface.}
\label{tab:masseystars}
\tablesize{\small}
\begin{tabular}{ccc c cc}
\toprule
\multirow{2}{*}{\textbf{Name}}      &\multicolumn{2}{c}{\textbf{Coordinates}} & \textbf{Spectral}    &   \boldmath{$V$}   & \multirow{2}{*}{\boldmath{$(B-V)$}}     \\\cmidrule{2-3}\cmidrule{5-5}
&\boldmath{$\alpha$}      & \boldmath{$\delta$}             & \textbf{Class}       &  \textbf{[mag]}   &             \\ \midrule
J013449.49+304127.2  &   01:34:49.46    & +30:41:27.1    & YSG:              & 16.468  & 0.854  \\
J013453.20+304242.8  &   01:34:53.17    & +30:42:42.7    & G/KV              & 17.031  & 0.906  \\
J013453.97+304043.4  &   01:34:53.94    & +30:40:43.3    & RSG:              & 19.492  & 1.558  \\
J013454.31+304109.8  &   01:34:54.28    & +30:41:09.7    & RSG               & 18.450  & 2.045  \\
J013455.06+304114.4  &   01:34:55.03    & +30:41:14.3    & B0I+Neb           & 18.246  & $-$0.103 \\
J013457.20+304146.1  &   01:34:57.17    & +30:41:46.0    & B3I               & 18.872  & $-$0.088 \\
J013458.77+304151.7  &   01:34:58.74    & +30:41:51.6    & RSG               & 19.121  & 1.605  \\
J013458.89+304129.0  &   01:34:58.86    & +30:41:28.9    & WC4               & 20.662  & 0.238  \\
J013459.07+304154.9  &   01:34:59.04    & +30:41:54.8    & RSG               & 19.030  & 1.986  \\
J013459.08+304142.8  &   01:34:59.05    & +30:41:42.7    & B2I:              & 19.306  & $-$0.163 \\
J013459.29+304128.0  &   01:34:59.26    & +30:41:27.9    & B0.5:I            & 18.822  & $-$0.112 \\
J013459.39+304201.2  &   01:34:59.36    & +30:42:01.1    & O8Iaf $^{a}$             & 18.254  & $-$0.142 \\
{\bf J013459.81+304156.9} &  {\bf 01:34:59.78}    & {\bf +30:41:56.8}    & {\bf RSG:}              & {\bf 19.156}  & {\bf 1.531}  \\
J013500.30+304150.9  &   01:35:00.27    & +30:41:50.8    & cLBV $^{b}$        & 19.298  & $-$0.073 \\
J013500.32+304147.3  &   01:35:00.29    & +30:41:47.2    & B0-2I             & 20.995  & $-$0.183 \\
J013501.36+304149.6  &   01:35:01.33    & +30:41:49.5    & Late O/Early B    & 19.346  & $-$0.279 \\
J013501.71+304159.2  &   01:35:01.68    & +30:41:59.1    & B1.5Ia            & 18.076  & $-$0.099 \\
J013502.06+304034.2  &   01:35:02.03    & +30:40:34.1    & RSG               & 18.500  & 1.365  \\
J013502.30+304153.7  &   01:35:02.27    & +30:41:53.6    & B0.5Ia            & 18.933  & $-$0.099 \\
J013505.37+304114.9  &   01:35:05.34    & +30:41:14.8    & WC4-5             & 19.061  & $-$0.293 \\
{\bf  J013505.74+304101.9}  &  {\bf  01:35:05.71}    & {\bf +30:41:01.8}    & {\bf O6III(f)+Neb}      & {\bf 18.218}  & {\bf $-$0.207} \\
J013506.87+304149.8  &   01:35:06.84    & +30:41:49.7    & B0.5Ib            & 18.655  & $-$0.181 \\
{\bf   J013507.43+304132.6}  &  {\bf   01:35:07.40}    & {\bf  +30:41:32.5}    & {\bf  RSG}               & {\bf  18.582}  & {\bf  1.991}  \\
J013507.53+304208.4  &   01:35:07.50    & +30:42:08.3    & RSG               & 19.961  & 1.739  \\
\bottomrule
\end{tabular}
\begin{tabular}{ccc}
\multicolumn{1}{c}{\footnotesize $^{a}$ later classified as Of/late-WN by~Humphreys et al. \cite{Humphreys2017}; $^{b}$ later classified as a FeII emission-line star by~Humphreys et al. \cite{Humphreys2014}.}
\end{tabular}
\end{table}

The field around GR\,290 is not yet sufficiently explored as it consists  mostly of faint ($V>18$~mag) stars  that require large telescopes for acquiring the spectra. Fortunately, some of the surrounding stars happened to lay on the slit during the long-slit observations of the object, thus that analysis of such data may provide additional information on the stellar contents and interstellar extinction in the vicinity of Romano's star. Therefore, we retrieved from General observational archive of Special Astrophysical observatory of Russian Academy of Sciences (SAO RAS)\footnote{General observational archive of Special Astrophysical observatory is available at \url{https://www.sao.ru/oasis/cgi-bin/fetch?lang=en}.} all long-slit spectra of GR\,290 obtained on Russian 6-m telescope with the Spectral Camera with Optical Reducer for Photometric and Interferometric Observations (SCORPIO) \citep{AfanasievMoiseev2005} during the  years 2005--2016. We also utilised the spectra obtained with the OSIRIS spectrograph on the {\it Gran Telescopio Canarias (GTC)} and analysed  by Maryeva et al.~\cite{GR290AandA} and Maryeva et al.~\cite{GR290preparation}. We reduced these spectra in a  uniform way using the ScoRe package\footnote{ScoRE package available at \url{http://www.sao.ru/hq/ssl/maryeva/score.htm}.} initially created for the SCORPIO data reduction, and~extracted the spectra of all stars crossed by the slit. To~perform the spectral classification of these stars, we used an automatic code based on the  $\chi^{2}$ fitting with spectral standards from STELIB\footnote{STELIB is availabte at \url{http://webast.ast.obs-mip.fr/stelib}.} (see Le Borgne et al.~\cite{STELIB}) in the same way as used by  Maryeva et al.~\cite{MaryevaChentsov2016MNRAS}. The~stars with spectra extracted and analysed in this way are marked with green circles in Figure~\ref{fig_map}, and~their estimated spectral classes, measured positions and photometric magnitudes are listed in Table~\ref{tab:nearstars}.   The resulting spectra of the stars in flux units are shown in Figures~\ref{star1}--\ref{star5}.

\begin{table}[H]
\centering
\caption{The sample of stars in the field around GR\,290 studied in this work. N corresponds to the labels in Figure~\ref{fig_map}. $V$ and $(B-V)$ taken from Massey et al.~\cite{Massey2016ubv}.}
\label{tab:nearstars}
\tablesize{\small}
\begin{tabular}{lcc cc cc}
\toprule
\multirow{2}{*}{\textbf{N}}  &\multicolumn{2}{c}{\textbf{Coordinates}} & \textbf{Spectral}        &  \boldmath{$V$}   & \multirow{2}{*}{\boldmath{$(B-V)$}}   & \multirow{2}{*}{\textbf{Instrument}}        \\\cmidrule{2-3}\cmidrule{5-5}
&  \boldmath{$\alpha$}      &  \boldmath{$\delta$}            & \textbf{Class}           &  \textbf{[mag]}   &           &                            \\ \midrule
1 $^{a}$   & 01:34:59.79    &   +30:41:56.9          &  RSG / M0-M1             &   19.156 & 1.531     & OSIRIS                  \\
2         & 01:35:00.69    &   +30:42:07.5          &  RSG /  K3-K4            &   20.507 & 1.613     & OSIRIS                  \\
3         & 01:35:00.90    &   +30:40:18.2          &  RSG /  K5-M0            &          &           & SCORPIO                 \\
4         & 01:35:01.87    &   +30:41:57.3          &   B5-B7                  &   19.980 & $-$0.079    & OSIRIS                  \\
5         & 01:35:02.37    &   +30:43:32.1          &   F:                     &   21.773 & 0.059     & OSIRIS                  \\
6         & 01:35:03.29    &   +30:40:42.5          &  RSG /  K-M                    &   20.339 & 1.088     & SCORPIO                 \\
7         & 01:35:04.37    &   +30:42:53.1          &   G4-K1                  &   16.459 & 0.733     & SCORPIO                 \\
8 $^{b}$   & 01:35:05.76    &   +30:41:02.2          & star with em.lines       &          &           & SCORPIO                 \\
9         & 01:35:05.87    &   +30:41:14.2          & star with em.lines       &   18.522 & $-$0.082    & SCORPIO                  \\
10 $^{c}$  & 01:35:06.14    &   +30:41:29.0          & F4-F6 V                  &   17.365 & 0.561     & SCORPIO                  \\
11        & 01:35:07.09    &   +30:42:12.4          & F4-F7 V                  &   14.888 & 0.592     & SCORPIO                  \\
12 $^{d}$  & 01:35:07.15    &   +30:41:56.3          & G8-K1                    &   17.793 & 0.962     & OSIRIS                   \\
13 $^{e}$  & 01:35:07.40    &   +30:41:32.5          & RSG / M0-M1              &   18.582 & 1.991     & SCORPIO                  \\
14        & 01:35:09.63    &   +30:41:46.1          & A9-F0                    &   20.817 & 0.246     & OSIRIS  SCORPIO          \\
15        & 01:35:11.40    &   +30:42:50.8          & F4-G2 V                  &   20.164 & 0.647     & SCORPIO                  \\      
16        & 01:35:11.66    &   +30:44:08.3          & hot star                 &   20.490 & 0.161     & SCORPIO                  \\
17        & 01:35:12.05    &   +30:43:27.2          & cool star                &   20.149 & 1.220     & SCORPIO                  \\    
& 01:35:14.10    &   +30:44:23.3             & hot star with abs.lines  &   17.654 &$-$0.038    & SCORPIO \\ \bottomrule
\end{tabular}\\
\begin{tabular}{@{}c@{}}
\multicolumn{1}{p{\textwidth -.88in}}{\footnotesize $^{a,e}$ Stars classified as RSG by Massey et al. \cite{Massey2006}; $^{b}$ star classified as O6III(f)+Neb by Massey et al. \cite{Massey2006}; $^{c,d}$ stars classified as foreground objects by  Massey et al. \cite{Massey2016ubv}.}
\end{tabular}
\end{table}

As we can see in Figure~\ref{fig_map}, our sample of stars partially intersect with the ones studied by Massey~et~al.~\cite{Massey2006,Massey2016ubv}. We were able to refine the estimates of spectral classes for J013459.81+304156.9 and J013507.43+304132.6, classified earlier as just red supergiants (RSG)  \citep{Massey2006}, as~well as for J013506.17+304129.1 and J013507.18+304156.4 as foreground objects according to  \citet{Massey2016ubv}. Our sample contains three more RSGs, which were not previously reported, and~four hot stars, with~only one (J013505.74+304101.9 with O6III(f)+Neb spectral class) known before \citep{Massey2006}. Among~three others, the~spectrum of J013505.76+304102.21 displays the He~I emission and strong nebular lines. The~second one, J013514.1+304423.21, has a spectral slope corresponding to high temperature, and~shows H and He absorption lines, while the last, J013501.87+304157.3, was preliminary  classified as B5--B7~supergiant.

Knowing the spectral classes of these stars, and~therefore their intrinsic colour indices, allows us to estimate the interstellar extinction around GR\,290. Its value is comparable to the  galactic foreground extinction value of $E_{(B-V)} =0.052$ (according to the NED extinction calculator \citep{schlegel98}). We did not register any star with higher reddening in the vicinity of GR\,290.


\section{Photometry}\label{photometry}

Photometric observations of GR\,290 were initiated in the early 1960s by the Italian astronomer Giuliano Romano in the Asiago Observatory~\cite{romano}.
He obtained a light curve with the brightness of a star varying irregularly between 16$\magn{.}7$ and 18$\magn{.}1$, and~classified it as a variable of the Hubble-Sandage type based on the shape of the light curve and GR\,290's colour index.

Subsequent photometric investigations of GR\,290 were undertaken by Kurtev et al.~\cite{Kurtev2001} and later by Zharova et al.~\cite{Zharova2011}.    The cumulative light curve derived in the latter work and covering half a century shows that GR\,290 exhibits irregular light variations with different amplitudes and time scales~\citep{Zharova2011}. The~star shows large and  intricate wave-like variations, with~duration of the waves amounting to several years.      In general, its variability is irregular, with~the power spectrum fairly approximated by a red power-law spectrum~\citep{PashaLBV} (i.e., the one dominated by a long timescale variations).     Moreover,  Kurtev~et~al.~\cite{Kurtev2001} discovered short-timescale variability with amplitude $\sim$0$\magn{.}$5, which is also typical  an LBV~star.



\begin{figure}[H]
\centering
\includegraphics[width=1.0\linewidth,clip,trim=98 -5 0 0]{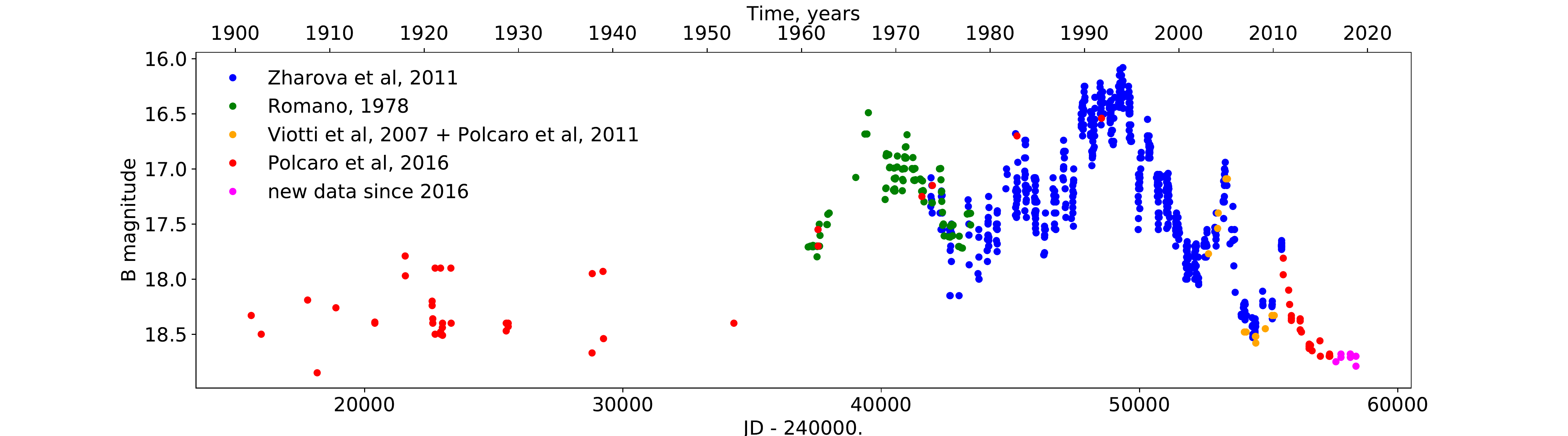}
\caption{The historical light curve of GR\,290 in the B-filter from 1901 to~2019. }
\label{fig_lc}
\end{figure}

Polcaro et al.~\cite{Polcaro2016} used various collections of photographic plates to further extend the historical light curve back to the beginning of the 20th century.  The~data between 1900 and 1950 suggest that no significant eruption took place during that half   century.  On~the contrary, after~1960, two clear, long-term eruptions are evident (see Figure~\ref{fig_lc}).

New photometric data, collected in Table~\ref{table_phot} and shown as   magenta dots in Figure~\ref{fig_lc}, confirm the conclusion of Maryeva et al.~\cite{GR290AandA} and Calabresi et al.~\cite{2014ATelCalabresi}  that the star has reached a long lasting visual minimum phase in 2013, and~its brightness has been relatively stable since then. 



\begin{table}[H]
\caption{New photometric observations of GR\,290 acquired by our group since Polcaro et al.~\cite{Polcaro2016}. \label{table_phot}}
\centering
\tablesize{\small}
\begin{tabular}{l lrlr lrlr l}
\toprule
\textbf{Date}    & \multicolumn{2}{c}{\boldmath{$B$}}& \multicolumn{2}{c}{\boldmath{$V$}}   & \multicolumn{2}{c}{\boldmath{$R$}}  & \multicolumn{2}{c}{\boldmath{$I$}}  & \textbf{Obs} \boldmath{$^{a}$} \\ \midrule
31  July 2016    &  18.75  &  0.03        &  18.77   &    0.04        &  18.59     & 0.04        &  18.66        &   0.05   &   Loiano                      \\
4 August 2016     &  18.75  &  0.03        &  18.77   &    0.04        &  18.59     & 0.04        &  18.66        &   0.05   &   Loiano                      \\
29 October  2016     &         &              &  18.60   &    0.1         &            &             &               &          &   ARA                      \\
30 October  2016     &         &              &  18.80   &                &            &             &               &          &   Loiano                      \\
31  October  2016    &         &              &          &                &  18.68     & 0.15        &               &          &   ARA                      \\
28  December  2016   &         &              &  18.78   &   0.15         &  18.71     & 0.15        &               &          &   ARA                      \\
16 February  2017     &  18.70  &  0.04        &  18.68   &   0.04         &  18.59     & 0.05        &  18.73        &  0.06    &   Loiano                      \\
28  July 2017    &         &              &  18.80   &   0.04         &  18.62     & 0.05        &               &          &   Loiano                      \\
30  July 2017    &         &              &  18.87   &   0.12         &            &             &               &          &   ARA                      \\
17 December 2017     &         &              &  18.67   &   0.10         &            &             &               &          &   ARA                      \\
14   February  2018     &  18.69  &  0.05        &  18.77   &   0.06         &  18.64     & 0.08        &               &          &   RTT-150                     \\
19  August 2018    &         &              & 18.81    &   0.10         &            &             &               &          &   ARA                      \\
4   September  2018    &  18.53  &  0.04        & 18.67    &   0.04         &  18.54     & 0.04        &  19.48        &  0.04    &   CMO                         \\
11  September  2018    &         &              & 18.84    &   0.04         &  18.64     & 0.04        &               &          &   Loiano                      \\
13  September  2018    &  18.65  &  0.04        & 18.73    &   0.04         &  18.63     & 0.04        &  19.51        &  0.04    &   CMO                         \\
19 September  2018     &  18.74  &  0.05        & 18.83    &   0.04         &  18.70     & 0.04        &               &          &   Loiano                      \\
10 January 2019    &         &              & 18.84    &   0.06         &  18.74     & 0.05        &               &          &   Loiano                      \\
\bottomrule
\end{tabular}\\
\begin{tabular}{@{}c@{}}
\multicolumn{1}{p{\textwidth -.88in}}{\footnotesize $^{a}$ observatories:  Loiano: 1.52\,m telescope at the Loiano station of the Bologna Astronomical; Observatory-INAF. ARA: 37 cm telescope of the Associazione Romana Astrofili at Frasso Sabino (Rieti); RTT-150: 1.5\,m Russian--Turkish telescope. CMO: 2.5\,m telescope of the Caucasian Mountain Observatory.}
\end{tabular}
\end{table}

\begin{figure}[H]
\centering
\includegraphics[width=1.0\linewidth]{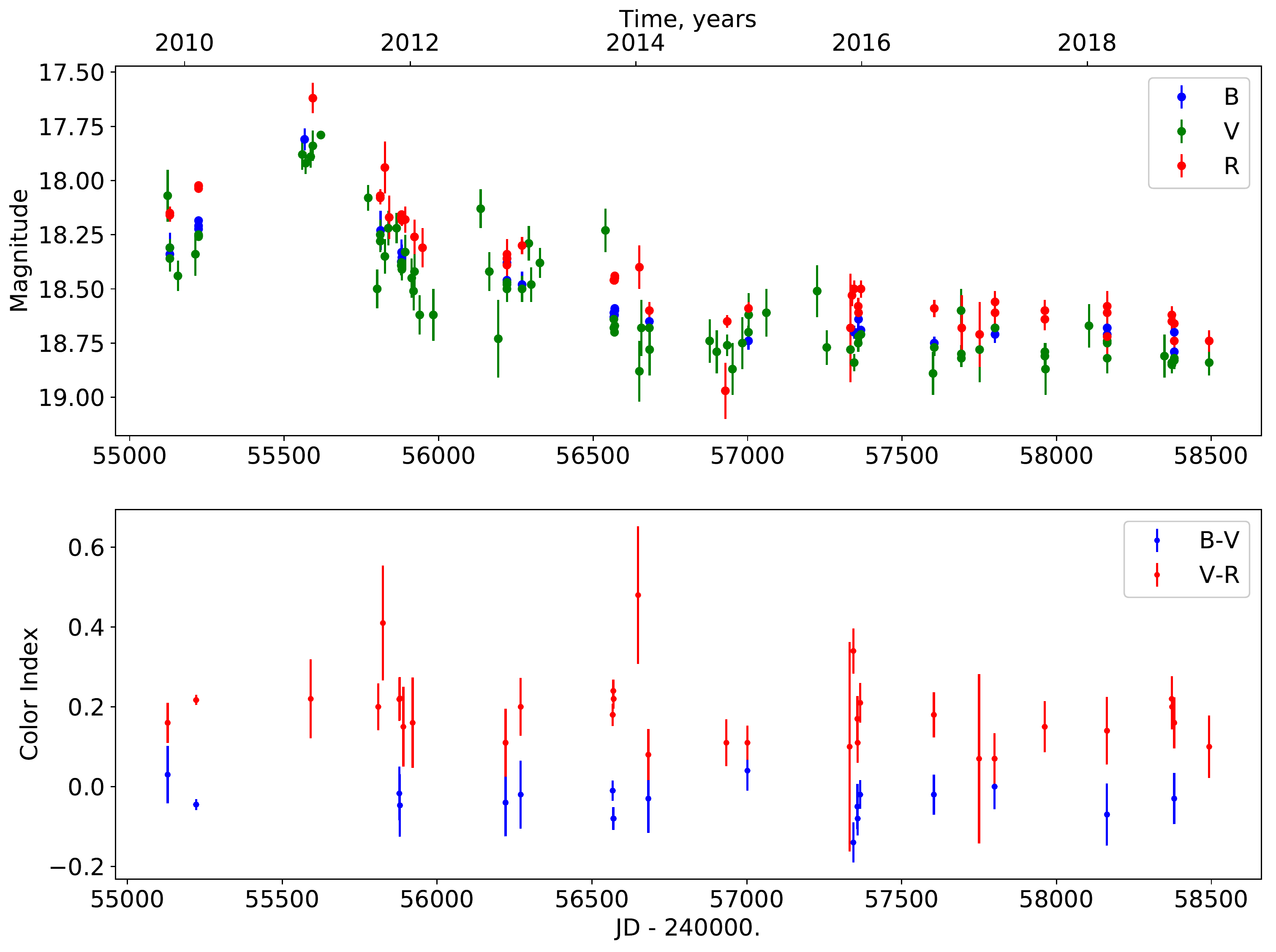}
\caption{(\textbf{top})  Light curve of GR\,290 in the B, V and R filters obtained by us between 2010 and 2019 and partially published in Polcaro et al.~\cite{Polcaro2016}.  (\textbf{bottom}) $(B-V)$ and $(V-R)$ colour indices for the same time interval.}
\label{fig_newphotometry}
\end{figure}

It is generally observed that, during~the S\,Dor { cycle}, the~colour of a typical LBV is bluer at the light minimum than close to the light maximum. In~contrast, Polcaro et al.~\cite{Polcaro2016} demonstrated that $(B-V)$ colour of Romano's star is constant over time, within~the error bars. There is no clear evidence for a variation of $(B-V)$ as a function of the visual magnitude, and~our new photometry obtained after 2015 confirms this conclusion (see Figure~\ref{fig_newphotometry}). This is consistent with Romano's star being hotter (about 30,000 K) than a typical LBV, with~the slope of optical spectrum defined by a Raleigh-Jeans power-law~tail.

\unskip


In other spectral ranges, the~object is much less studied than in optical range. Only a single measurement of its  magnitude is available in ultraviolet and infrared ranges, corresponding to different moments of time defined by a mean epoch of the individual survey (Table~\ref{iruv}).
\begin{table}[H]
\caption{Stellar magnitudes of GR\,290 in ultraviolet and infrared~range. \label{iruv}}
\centering
\begin{tabular}{lllllll}
\toprule
\multicolumn{2}{c}{\textbf{GALEX}}               &\multicolumn{3}{c}{\textbf{2MASS}}                    &\multicolumn{2}{c}{\textbf{Spitzer}}                \\ \cmidrule{1-7}
Far-UV 1516 \AA        & Near-UV 2267 \AA  &   J             &  H             &  K        &   3.6 $\upmu$m &  4.5 $\upmu$m                \\\cmidrule{1-7}
$17.692\pm0.031$      & $17.438\pm0.013$ & $16.834\pm0.128$&$16.702\pm0.292$& 16.657    &  16.335      & 15.921                     \\ \midrule
\multicolumn{2}{c}{\cite{Mudd2015UV}}   &\multicolumn{3}{c}{\cite{2MASS}}             &\multicolumn{2}{c}{\cite{Spitzer1}}        \\
\bottomrule
\end{tabular}
\end{table}





\section{Spectroscopy and Determination of Physical~Parameters}\label{spectroscopy}



The first description of optical spectrum of Romano's star can be found in the article of Humphreys~\cite{Humphreys1980}. The spectrum was obtained in August 1978  at Kitt Peak National Observatory, when brightness of the star was $V=18.00\pm0.02$ \citep{Humphreys1980}. Humphreys noted: ``Its spectrum shows emission lines of hydrogen and He\,{\sevensize I}. There are no emission lines of Fe\,{\sevensize II} or [Fe\,{\sevensize II}].'' and classified the star as a  peculiar emission-line object\footnote{In \citet{Humphreys1980}, Romano's star is identified as B\,601.} \citep{Humphreys1980}.

\begin{figure}[H]
\centering
\includegraphics[width=1.0\linewidth]{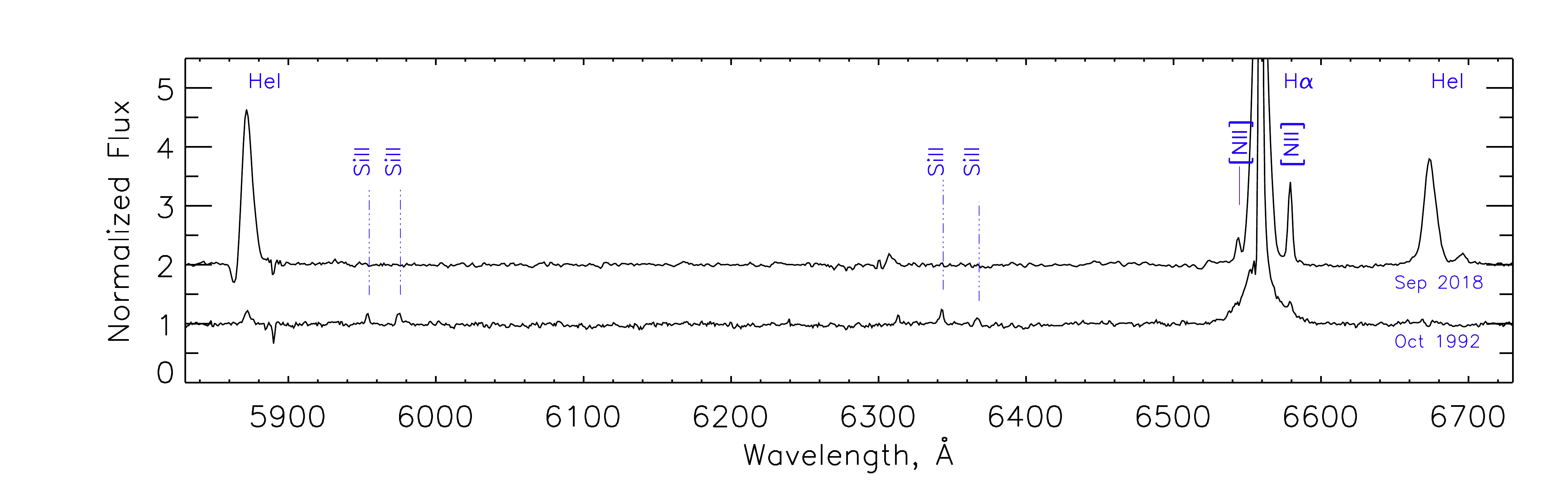}
\caption{Comparison of normalized optical spectra of GR\,290 obtained with Calar Alto/TWIN in October 1992 by \citet{szeifert}
and with GTC/OSIRIS in September 2018. Spectra are displaced vertically for illustrative purposes.}
\label{fig_twin}
\end{figure}

{ In 1992, T.~Szeifert obtained a spectrum of Romano's star right before the historical maximum of its brightness. }
\citet{szeifert} described it as ``Few metal lines are visible, although~a late B spectral type is most likely'' (Figure~\ref{fig_twin}). On~the other hand, Sholukhova et al.~\cite{Olga97} obtained the next spectrum in August 1994 and classified  the star as a WN star candidate.
Since 1998, regular observations of GR\,290 carried out on the Russian 6m \citep{fabrika,Sholukhova2011} and spectra published by Sholukhova et al.~\cite{Sholukhova2011} indicate that the spectrum of GR\,290 has not reverted to a B-type spectrum.  Thus, Szeifert's~\cite{szeifert} spectrum is unique and corresponds to the coldest and  brightest state of the star measured so far. 

Studies of GR\,290 devoted to its spectral variability show that its spectral type changes between WN11 and WN8 \citep{maryeva2010,polcaro10,Sholukhova2011}. Since the beginning of the 2000s,  it has made this transition twice \citep{Polcaro2016}. Viotti~et~al.~\cite{Viotti2006, Viotti2007} first described an anticorrelation between equivalent width of 4600--4700 \AA\ blend and the brightness. Later, \citet{maryeva2010} found a correlation of spectral changes and the visual brightness typical for LBVs: the brighter it is, the~cooler the spectral type. However, as~noted by Humphreys et al.~\cite{Humphreys2014}, GR\,290 does not exhibit S\,Dor like transitions to the cool state with an optically thick wind, but~instead varies between two hot states characterised by WN spectroscopic features. Among~all known LBVs, only HD\,5980 \citep{Georgiev2011} convincingly shows a hotter spectrum in the minimum of brightness.
Other LBV stars showing WN-like spectrum in quiescent ``hot'' phase usually stop at colder spectral types  such as  WN11 (for example AG\,Car \citep{Groh2009AGCar} and WS\,1 \citep{KniazevGvaramadze2015WS1,Gvaramadze2012}) or  Ofpe/WN9 (for example R\,127~\cite{Walborn2008R127} and HD\,269582~\cite{Walborn2017LBV}).

As already mentioned, since the autumn of 2013, GR\,290 is in a minimum brightness state with $V$~=~18.7--18.8~mag. Due to this, it has been challenging to obtain its spectra with good enough quality for wind speeds to be adequately estimated. In~summer of 2016, GR\,290 was observed with the Optical System for Imaging and low-Intermediate-Resolution Integrated Spectroscopy (OSIRIS) on the {\it Gran Telescopio Canarias (GTC)} \citep{GR290AandA}. These observations gave the best spectral resolutions and signal-to-noise ratios ever obtained for this object, and~allowed to estimate an average radial velocity (RV) of the object, RV(GR\,290) = $-$163 $\pm$ 32~km s$^{-1}$, which is consistent, within~the uncertainties, with~the heliocentric velocity $-$179 $\pm$ 3~km s$^{-1}$ of M\,33~galaxy.

\begin{figure}[H]
\centering
\includegraphics[width=1.0\linewidth]{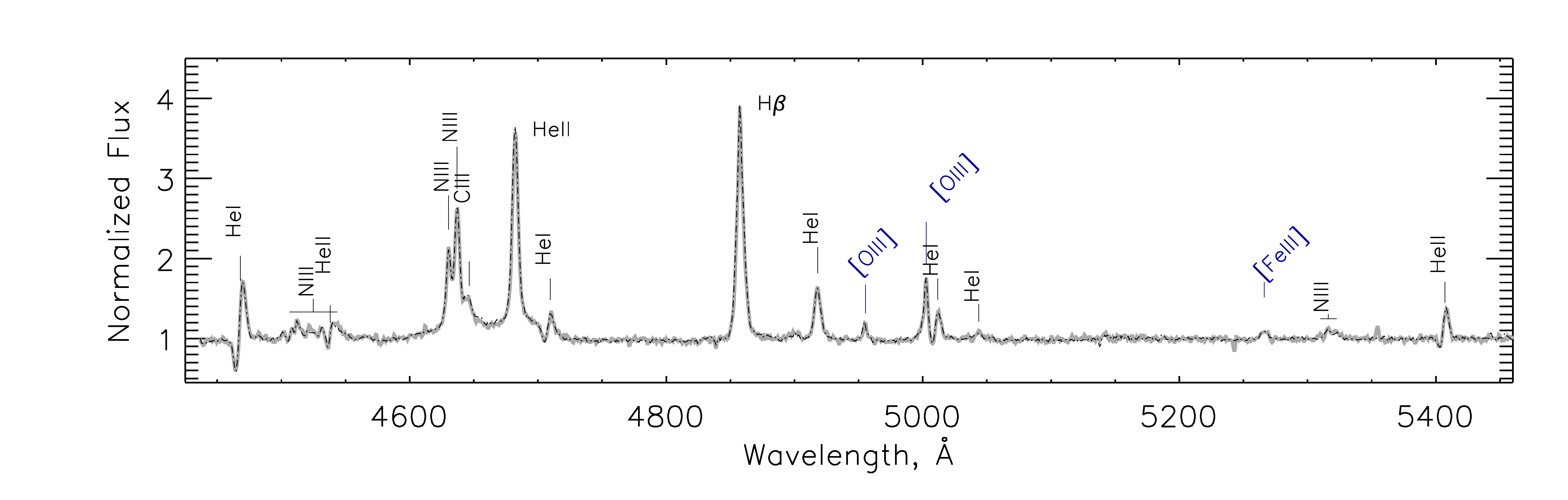}
\caption{Comparison of normalised optical spectra of GR\,290 obtained with GTC/OSIRIS in July 2016 (grey thick line)
and September 2018 (black dash-dotted line). Spectra are nearly identical.}
\label{fig_gtc1618}
\end{figure}

New spectra of GR\,290 were obtained with the OSIRIS spectrograph in September 2018 \citep{GR290preparation}. Detailed analysis of the spectra obtained in 2016 and 2018 did not reveal any changes (Figure~\ref{fig_gtc1618}). As~before, the~star displays a WN8h spectrum with forbidden nebular~lines.

\begin{figure}[H]
\centering
\includegraphics[width=0.85\linewidth,clip,trim=35 10 0 0]{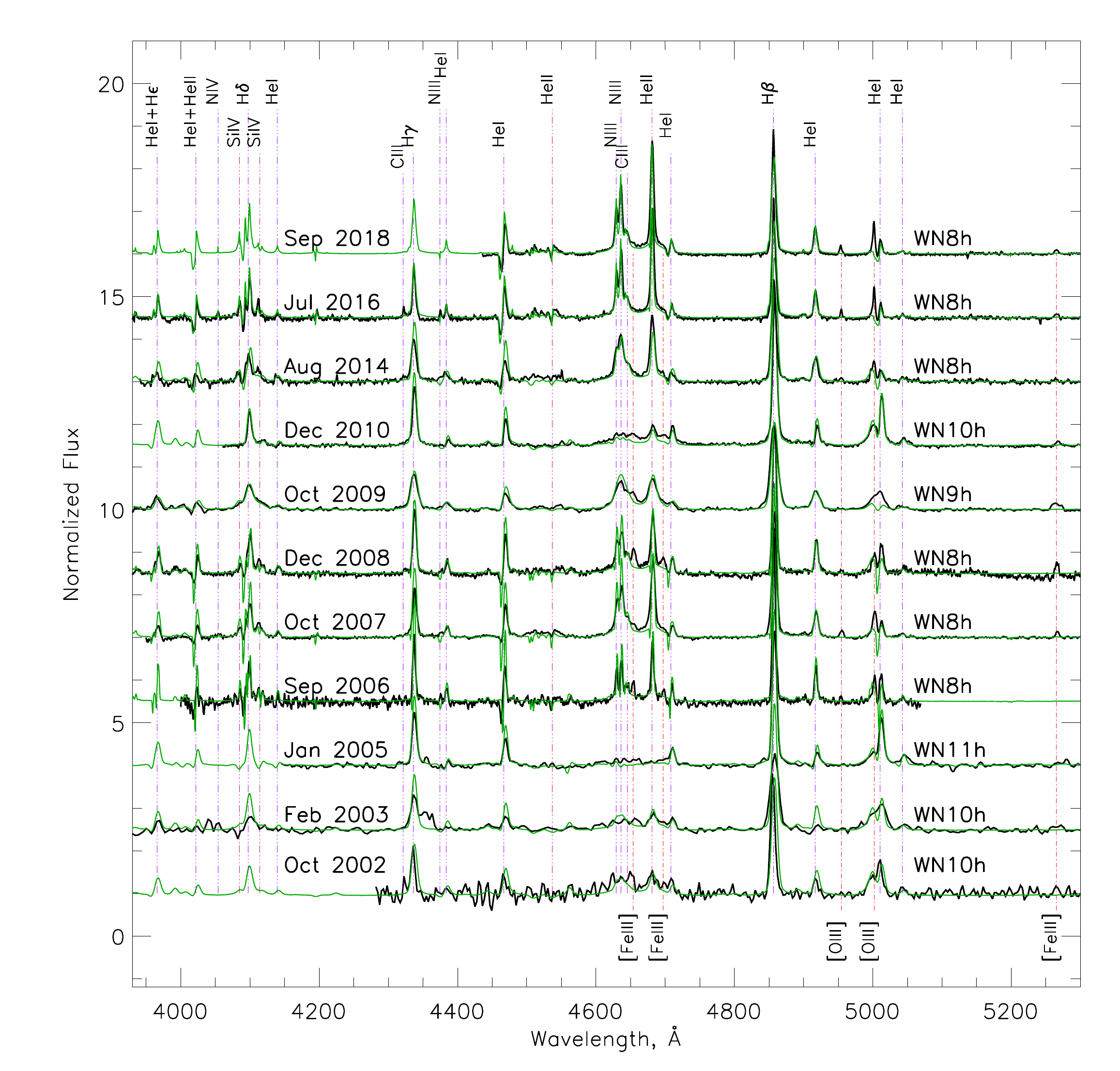}
\caption{Normalised optical spectra of GR\,290 compared with the best-fit CMFGEN models (green line). The~model spectra are convolved with a Gaussian
instrumental profile. Description of observational data may be found in~\cite{Polcaro2016,GR290AandA,GR290preparation}.
Notice that ``September 2006'' spectrum was obtained by P.~Massey with WIYN 3.5\,m telescope \citep{Massey2007cLBV}. Spectral types are estimated based primarily on relative strengths of N\,{\sevensize V}, N\,{\sevensize IV}, N\,{\sevensize III}, N\,{\sevensize II} and He\,{\sevensize II}\,$\lambda$4686 emission lines \citep{smithprinja}.
Spectra are displaced vertically for illustrative~purposes.}
\label{fig_modelsp}
\end{figure}

The large number of acquired spectra allows   tracking  the quantitative changes of physical parameters of the star over time.  To do it, a~numerical modeling of GR\,290's atmosphere using CMFGEN code \citep{Hillier5} was started by \citet{maryeva2012}, who constructed models for two states---the luminosity maximum of 2005 and the minimum of brightness in 2008. Then, Clark~et~al.~\cite{ClarkLBV2012} estimated the parameters of GR\,290 during the moderate luminosity maximum of 2010. Polcaro~et~al.~\cite{Polcaro2016} built nine models for the most representative spectra acquired between 2002 and 2014.  The~results of calculations from Polcaro et al.~\cite{Polcaro2016},  Clark~et~al.~\cite{ClarkLBV2012} and Maryeva et al.~\cite{GR290AandA} are summarised in Table~\ref{tbl_parmodel}, along with the parameters estimated using the spectrum of September 2018. Comparisons of observed spectra with corresponding models are shown in Figure~\ref{fig_modelsp}.
\begin{table}[H]
\centering
\caption{Derived properties of Romano's star at the moments corresponding to different acquired spectra. H/He indicates the hydrogen number fraction relative to helium,  $f$ is the filling factor of the stellar wind. Details of modeling may be found in~\cite{Polcaro2016,ClarkLBV2012,GR290AandA}. }
\label{tbl_parmodel}
\scalebox{0.86}[0.86]{
\begin{tabular}{lll l lll llc lcc}
\toprule
\multirow{2}{*}{\textbf{Date}}           & \textbf{V }     &   \textbf{Sp.}   &\boldmath{$T$}$_{\textbf{eff}}$& \multirow{2}{*}{\textbf{log} \boldmath{$T$}$_{\textbf{eff}}$} & \boldmath{$R_{2/3}$      }  & \boldmath{$L_*10^5$}    & \textbf{log} \boldmath{$L_*$}      &   \boldmath{$\dot{M}_{cl}10^{-5}$}  &  \multirow{2}{*}{\textbf{f}}        & \boldmath{$v_{\infty}$}       & \multirow{2}{*}{\textbf{H/He}}    & \multirow{2}{*}{\textbf{Ref.}}  \\
&\textbf{[mag]}   &  \textbf{typ}e   &   \textbf{[kK]}     &                     &\boldmath{[$\rm R_{\odot}$]} & \boldmath{[$L_{\odot}$]}&\boldmath{[$\rm L_{\odot}$]}&\boldmath{[$ \rm M_{\odot}/yr$]}&          &\boldmath{[$\rm km/s$]}  &         &   \\  \midrule

Oct. 2002      & 17.98  & WN10h   & 28.1        &  4.45               &  31.6            & 5.6          &  5.75           &    1.9                    & 0.15     & $250 \pm 100$      & 1.7     & \cite{Polcaro2016}  \\
Feb. 2003      & 17.70  & WN10.5h & 28.0        &  4.45               &  37              & 7.5          &  5.875          &    2.2                    & 0.15     & $250 \pm 50$       & 1.7     & \cite{Polcaro2016}  \\
Jan. 2005      & 17.24  & WN11h   & 23.6        &  4.37               &  54              & 8.2          &  5.91    &    3.5                    & 0.15     & $250 \pm 50$       & 1.7     & \cite{Polcaro2016}  \\
Sep. 2006      & 18.4   & WN8h    & 30.7        &  4.49               &  24              & 4.6          &  5.66           &    1.3                    & 0.15     & $250 \pm 100$      & 1.7     & \cite{Polcaro2016}  \\
Oct. 2007      & 18.6   & WN8h    & 33.5        &  4.53               &  20              & 4.5          &  5.65           &    1.55                   & 0.15     & $370 \pm 50$       & 1.7     & \cite{Polcaro2016} \\
Dec. 2008      & 18.31  & WN8h    & 31.6        &  4.50               &  23.5            & 5.0          &  5.7            &    1.9                    & 0.15     & $370 \pm 50$       & 1.7     & \cite{Polcaro2016}  \\
Oct. 2009      & 18.36  & WN9h    & 31.6        &  4.50               &  23.8            & 5.1          &  5.7            &    1.7                    & 0.15     & $300 \pm 100$      & 1.7     & \cite{Polcaro2016}  \\
Sep. 2010 $^{a}$&17.8    & WN10h   &  26         &  4.41               &  41.5            &              &   5.85          &   2.18                    & 0.25     &        265         &1.5      & \cite{ClarkLBV2012}  \\
Dec. 2010      & 17.95  & WN10h   &  26.9       &  4.43               &  33              & 5.3          &  5.72           &    2.05                   & 0.15     & $250 \pm 100$      & 1.7     & \cite{Polcaro2016}  \\
Aug. 2014      & 18.74  & WN8h    &  32.8       &  4.52               &  19              & 3.7          &  5.57           &    1.4                    & 0.15     & $400 \pm 100$      & 1.7     & \cite{Polcaro2016}  \\
Jul. 2016      & 18.77  & WN8h    & 30.0        &  4.48               &  21              & 3.7          &  5.57           &    1.5                    & 0.15     & $620 \pm 50$       &  2.2    & \cite{GR290AandA}  \\
Sep. 2018      & 18.77  & WN8h    & 30.0        &  4.48               &  21              & 3.7          &  5.57           &    1.5                    & 0.15     & $620 \pm 50$       &  2.2    & \\ \bottomrule
\end{tabular}}\\
\begin{tabular}{ccc}
\multicolumn{1}{c}{\footnotesize $^{a}$ Clark et al.~\cite{ClarkLBV2012} assumed a distance to M\,33 of 964\,kpc.}
\end{tabular}
\end{table}

\begin{figure}[H]
\centering
\includegraphics[width=0.85\linewidth]{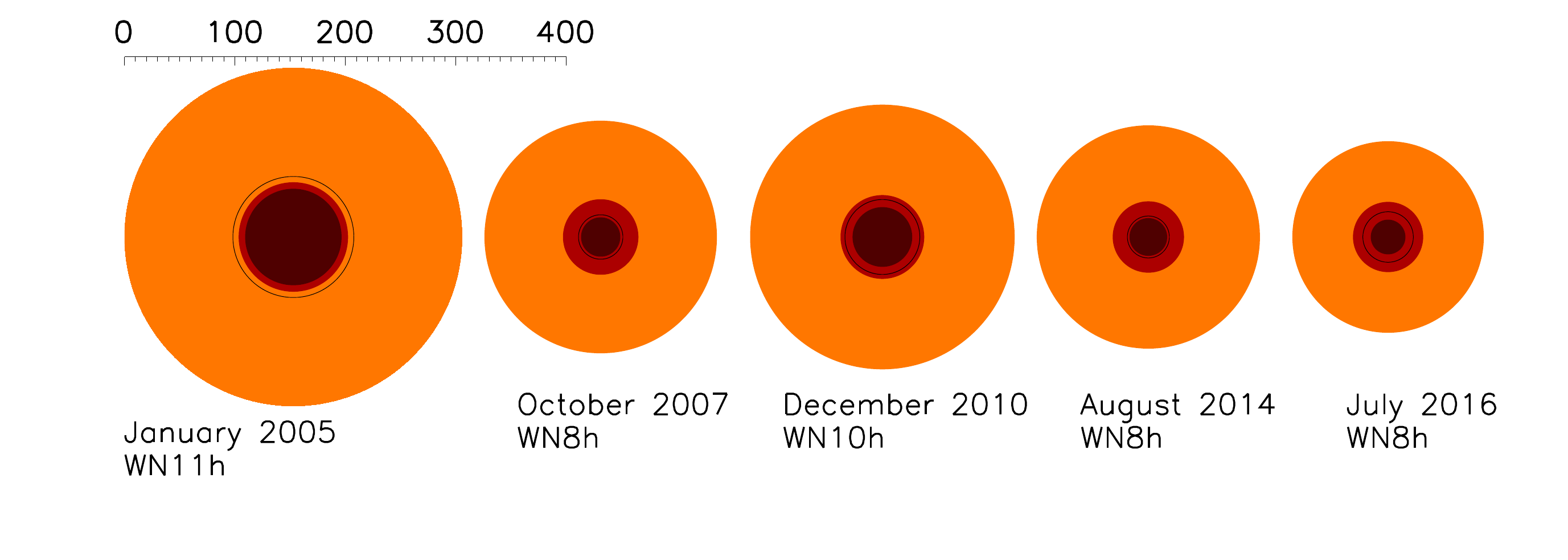}
\caption{Change of the wind structure and extent over time. The~region where $n_e\ge 10^{12}~\text{cm}^{-3}$ is shown in dark red, $10^{12}\ge n_e\ge 10^{11}~\text{cm}^{-3}$  in red, and  $10^{11}\ge n_e\ge 10^{10}~\text{cm}^{-3}$ in orange. 
Solid black line shows the radius where Rosseland optical depth ($\tau$) is 2/3. Scale in units $\Rsun$ is shown at the top.}
\label{fig_circ}
\end{figure}

Numerical calculations show that the bolometric luminosity of GR\,290 is variable, being higher during the phases of greater optical brightness~\cite{maryeva2012,Polcaro2016}.  At~the same time, the~ wind structure of GR\,290 also varies in correlation with brightness changes---the slow and dense wind at brightness maxima becomes faster and thinner at minima (Figure~\ref{fig_circ}), and~the effective temperature\footnote{Effective temperature is defined as a temperature at radius $R_{2/3}$, where the Rosseland optical depth is equal to 2/3.} of the star increases from 25 kK (with WN11h  spectral type) during the maximum of 2005 year to 31--33 kK (WN8h) during the~minima.



\begin{figure}[H]
\centering
\includegraphics[width=0.85\linewidth,clip,trim=25 -5 0 0]{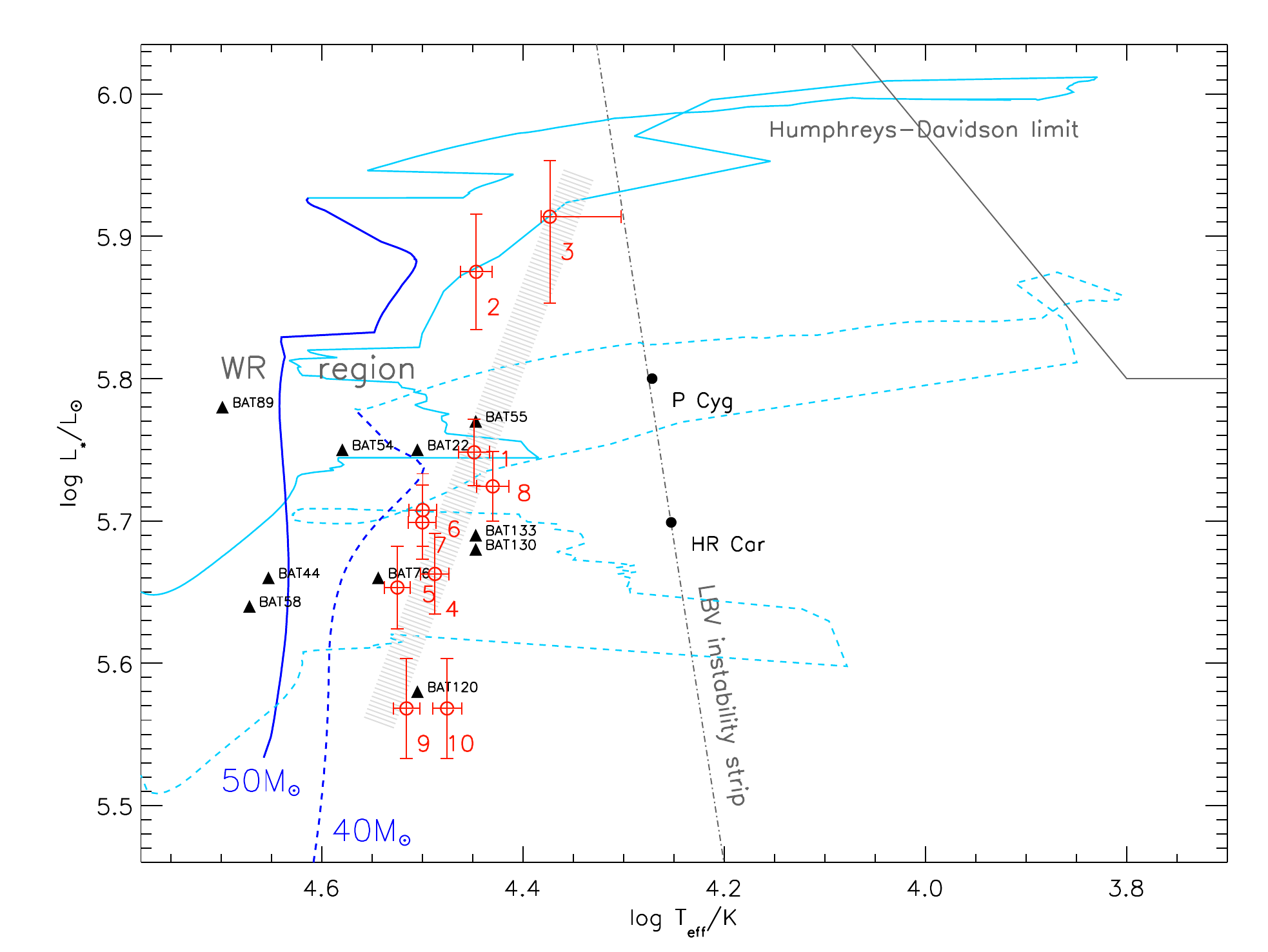}
\caption{Position of GR\,290 in the Hertzsprung-Russell diagram at different times. Numbers correspond to:   (1)  October 2002;   (2)  February 2003;   (3)  January 2005;
(4)  September 2006;   (5)  October~2007;   (6)  December 2008;   (7)  October 2009;   (8)  December 2010;   (9)  August 2014; and  (10)  July~2016 and September 2018.
The hatched strip shows an average line along which GR 290 moved during its recent luminosity cycles.
The Geneva tracks~\cite{Ekstrom} for $40~\Msun$ (dashed line) and $50~\Msun$ (solid line) with rotation are shown by blue lines,
with dark blue part corresponding to the hydrogen burning in stellar core.
Triangles mark the positions of late-WN stars from Large Magellanic Cloud (LMC), whose data were taken from Hainich et al.~\cite{HainichWNLMC}. In addition, the~positions of LBV stars P\,Cygni and HR\,Car are
shown with circles. Data for these objects were taken from the works of  \citet{NajarroPCyg} and  Groh et al.~\cite{grohHR}.
Grey solid line is Humphreys-Davidson limit~\cite{HumphreysDavidson}, grey dash-dotted line is LBV minimum instability strip as defined in~\cite{grohHR}.}
\label{fig_hrdiagram} 
\end{figure}

Figure~\ref{fig_hrdiagram} shows the positions of the star in the H-R diagram at different times. The~object clearly moves well outside the typical LBV instability strip { \citep{Wolf1989,Clark2005}}, deep inside the region of Wolf-Rayet stars, except~for a moment of maximum brightness in 2005.
On average, GR\,290 lays on the  40--50~$\Msun$  evolutionary tracks from the Geneva models~\cite{Ekstrom} with rotation.  Using CMFGEN, we found that hydrogen mass fraction in the atmosphere of  GR\,290 is 35\% \citep{GR290AandA}, and~used this estimation for determination of current stellar mass and age.   According to this tracks, the~Romano's star should now be 4.5--5.7~Myr old and should have  a mass of 27--38~$\Msun$.

\unskip

\section{Nebula}\label{nebula}

The presence of forbidden lines [N\,{\sevensize II}] 6548, 6584; [O\,{\sevensize III}] 4959, 5007; [Fe\,{\sevensize III}] 4658, 4701, 5270 and [Ar\,{\sevensize III}] 7136 in the spectrum of GR\,290 indicates that it has a nebula, but~it is not resolved in direct imaging because of the large distance to M33.
In 2005, Fabrika et al.~\cite{fabrika} first attempted to detect the nebula and study its spatial structure using the panoramic (3D) spectroscopic data acquired on Russian 6-m telescope, and~ reported the discovery of an extended structure in the velocity field of H$\beta$ line, with~an angular extent of $\sim 9\arcsec$ ($\sim 30$~pc) in the NE-SW direction.
An excess corresponding to the dust circumstellar envelope around the object has not been detected in the infrared (IR) emission \citep{Humphreys2014}.

Maryeva et al.~\cite{GR290AandA} performed a modeling of circumstellar nebula using using CLOUDY photoionisation code \citep{Ferland1998,Ferland2017} and the spectrum acquired on GTC/OSIRIS in order to reproduce the observed nebular emission lines that are clearly seen in the spectrum. GR\,290 was found to be surrounded by an unresolved compact H\,{\sevensize II} region with a most probable outer radius R = 0.8~pc and a hydrogen density $n_H$ = 160 cm$^{-3}$, and~having chemical abundances that are consistent with those derived from the stellar wind lines. Hence, this compact H\,{\sevensize II} region appears to be largely composed of material ejected from the~star.

In addition, the~recent analysis of the 2D spectra obtained with GTC/OSIRIS in September 2018 perpendicular to the dispersion at H$\alpha$ line indicates that 
the nebula has extended and asymmetric structure \citep{GR290preparation}. Its size is about 25--30~pc, similar to    typical H\,{\sevensize II} regions around O-stars. Based on the similarity of sizes and evolutionary status of GR\,290, we speculate that this extended nebula consists of   material ejected during O-supergiant~phase.

\section{Conclusions}\label{conclusion}

GR\,290 is is located in the outer spiral arm of the M\,33 galaxy at a projected distance of about 4~kpc from the centre. Its spatial location, the~proximity to the OB\,88 and OB\,89 associations, and~the similarity of their ages (about 4--5~Myr) as well as a basic concept that a large fraction of all stars, including massive stars, forms in clusters suggest the common origin of GR\,290 and OB\,89. It is tempting to suggest that GR\,290 may have escaped from the~association.

The evolution of LBVs during the S\,Dor cycles seems to occur in most cases roughly at constant bolometric luminosity (see,    e.g.,~\cite{HumphreysDavidson,dekoter96}).
However, a~decrease of bolometric luminosity from minimum towards the light maximum of the S\,Dor cycle were observed for several LBVs (e.g., S\,Dor~\cite{Lamers95} and AG\,Car~\cite{Groh2009AGCar}). \citet{Lamers95} interpreted it in terms of the radiative power being partially transformed into mechanical power in order to expand the outer layers of the star from minimum to maximum. In~contrast, spectral monitoring of  Romano's star during its recent peaks of activity, and~the numerical simulation of its stellar atmosphere based on acquired spectra, demonstrated that its bolometric luminosity varies in correlation with its visual brightness, i.e.,~$L_{\rm bol}$ increases   during its visual luminosity maxima \citep{Polcaro2016}.  \citet{Guzik2012} discussed several mechanisms that could trigger the large outburst activity and variations in bolometric magnitude as observed in GR\,290. An~interesting possibility is that the interplay between pulsations and rotational mixing lead to an unstable transport of H-rich material to the nuclear burning core.  In~this context, GR\,290 may be the ideal object for testing such theories. 

The star is hotter than most other LBVs { (Table~\ref{tablbvstars})}, and~lays outside of the LBV instability strip in the H-R diagram. On~the other hand, the~hydrogen abundance of the envelope appears higher than in late type WN stars, and~therefore, from the evolutionary and structural point of view, GR\,290 is less evolved than WN8h stars \citep{Polcaro2016}. This suggests that Romano's star may be a post-LBV object, the~transition phase between LBVs and Wolf-Rayet~stars.

The century long light curve of Romano's star shows that until the 1960s the object was in a long lasting quasi-stationary state,
a state to which it has returned in 2013, and~since then displaying a WN8h spectrum. While the spectral type during the early ``low'' state (pre-1960) is unknown,
from the observed correlation between the visual magnitude and spectral type, we may suggest that it also was WN8h.
The Galactic WN8 stars are known to be significantly more variable than the WRs with hotter spectral types~\cite{Moffat1989}.
Thus, it is tempting to speculate on the possibility that, in~analogy with GR\,290, other WN8s may have just recently passed through the LBV phase.
Hence, a~systematic investigation of archival data and constructing century long light curves for WN8-WN9 stars using archival photographic plates
will probably be able to uncover more objects similar to Romano's~star.

\begin{table}[H]
\centering
\caption{Comparison of Romano's star with other LBVs and LBV candidates which show WR like~spectra.}
\label{tablbvstars}
\scalebox{0.9}[0.9]{
\begin{tabular}{llcl}
\toprule
\multicolumn{1}{c}{\textbf{Star}} &   \multicolumn{1}{c}{\textbf{Sp.type}}  &  \multicolumn{1}{c}{\textbf{Ref.}}  &    \multicolumn{1}{c}{\textbf{Comments}}     \\
\midrule
Wray 15-751              &   O9.5 I   &  \citep{Sterken2008}                             &  LBVc,   Milky Way               \\

{Sk$-69^\circ\,279$}       &    {O9.2 Iaf} &{\citep{Gvaramadze2018Sk69}}       &   ex-/dormant LBV \citep{vanGenderen2001}, BSG evolved off  the Main Sequence \citep{Gvaramadze2018Sk69} \\


%
{Hen\,3-519}              &    {WN11}     &  {\citep{Toala2015}}                &   ex-/dormant LBV \citep{vanGenderen2001}, there are no    significant changes of brightness \citep{Davidson1993}\\


AG\,Car                  &   WN11     & \citep{Groh2009AGCar}                            &   LBV, Milky Way                 \\
WS\,1                    &   WN11     & \citep{KniazevGvaramadze2015WS1,Gvaramadze2012}  &   LBV, Milky Way                 \\
R\,127                   &   Ofpe/WN9 & \citep{Walborn2008R127}                          &   LBV, Large Magellanic Cloud    \\
HD\,269582               &   Ofpe/WN9 & \citep{Walborn2017LBV}                           &   LBV, Large Magellanic Cloud    \\
GR\,290                  &   WN8      & \citep{Polcaro2016,GR290AandA}                   &   post-LBV, M33                   \\
HD\,5980                &LBV+WN4+OI  & \citep{Foellmi2008,Koenigsberger2014}            & LBV, Small Magellanic Cloud    \\ \bottomrule
\end{tabular}
}
\end{table}

\vspace{6pt}

\authorcontributions{R.F.V., spectral analysis; M.C., C.R. and R.G., photometric monitoring and reduction of photometry data;  and O.M., numerical modeling of stellar atmosphere and reduced the spectroscopic material and manuscript preparation. G.K. was PI of the 2016 and 2018 GranTeCan observations and performed spectral analysis.    All authors discussed the results and commented on the~manuscript.}

\funding{This research was funded by CONACYT grant 252499, UNAM/PAPIIT grant IN103619,  Russian Foundation for Basic Research grant 19-02-00779  and Czech Science Foundation grant GA18-05665S. This project   received funding from the European Union's Framework Programme for Research and Innovation Horizon 2020 (2014-2020) under the Marie Sk\l{}odowska-Curie Grant Agreement No. 823734.}

\acknowledgments{We express our enormous gratitude to V.F.~Polcaro, who recently passed away, for~having stimulated our interest and studies of this unique~object. We thank Roman Zhuchkov, Oleg Egorov and Olga Vozyakova for obtaining the photometric observations on 1.5\,m Russian--Turkish telescope and 2.5\,m telescope of the Caucasian Mountain Observatory. 
We thank Thomas Szeifert and Philip Massey for the spectra obtained with Calar Alto/TWIN spectrograph in 1992 and with WIYN 3.5\,m telescope in September 2006.
We thank the GTC observatory staff for obtaining the spectra and Antonio Cabrera-Lavers for guidance in processing the~observations.
We thank Guest Editor Prof. Roberta M. Humphreys and our  anonymous referees for providing helpful comments and suggestions.
In this paper, we use data taken from the public archive of the SAO RAS. The work is partially based on the observation at 2.5-m CMO telescope that is supported by M.V. Lomonosov Moscow State University Program of Development. }

\conflictsofinterest{The authors declare no conflict of~interest.}


\abbreviations{The following abbreviations are used in this manuscript:\\

\noindent
\begin{tabular}{@{}ll}
CMO     & 2.5\,m telescope of the Caucasian Mountain Observatory of the Sternberg Astronomical Institute \\
&  of Moscow State University \\
GTC     & Gran Telescopio Canarias \\
LBV     & Luminous blue variable \\
BSG     & Blue supergiant \\
RSG     & Red supergiant \\
SAO RAS & Special Astrophysical observatory of Russian Academy of Sciences \\
WR      & Wolf-Rayet \\
WN      & Nitrogen-rich Wolf-Rayet stars \\
WC      & Carbon-rich Wolf-Rayet stars \\
YHG     & Yellow hypergiant \\
\end{tabular}}

\appendixtitles{yes} 
\appendix
\section{Spectra of Stars in Vicinity of GR\,290}
\unskip
\begin{figure}[H]
\includegraphics[width=1.0\linewidth,clip,trim=30 5 0 0]{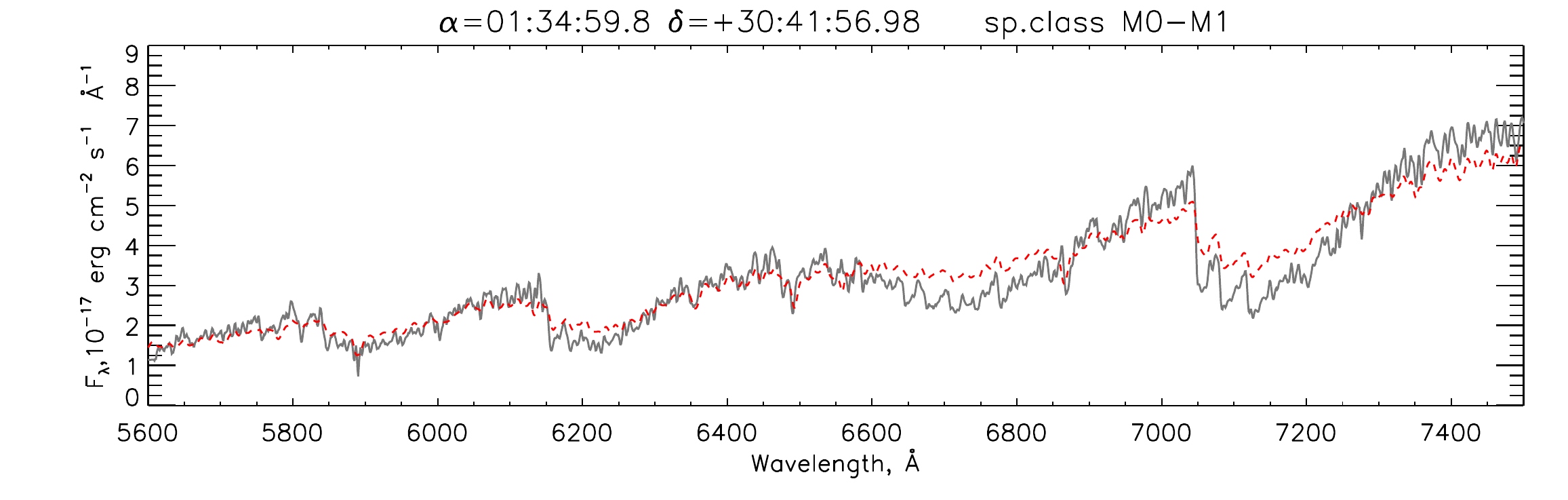}\\
\includegraphics[width=1.0\linewidth,clip,trim=30 5 0 0]{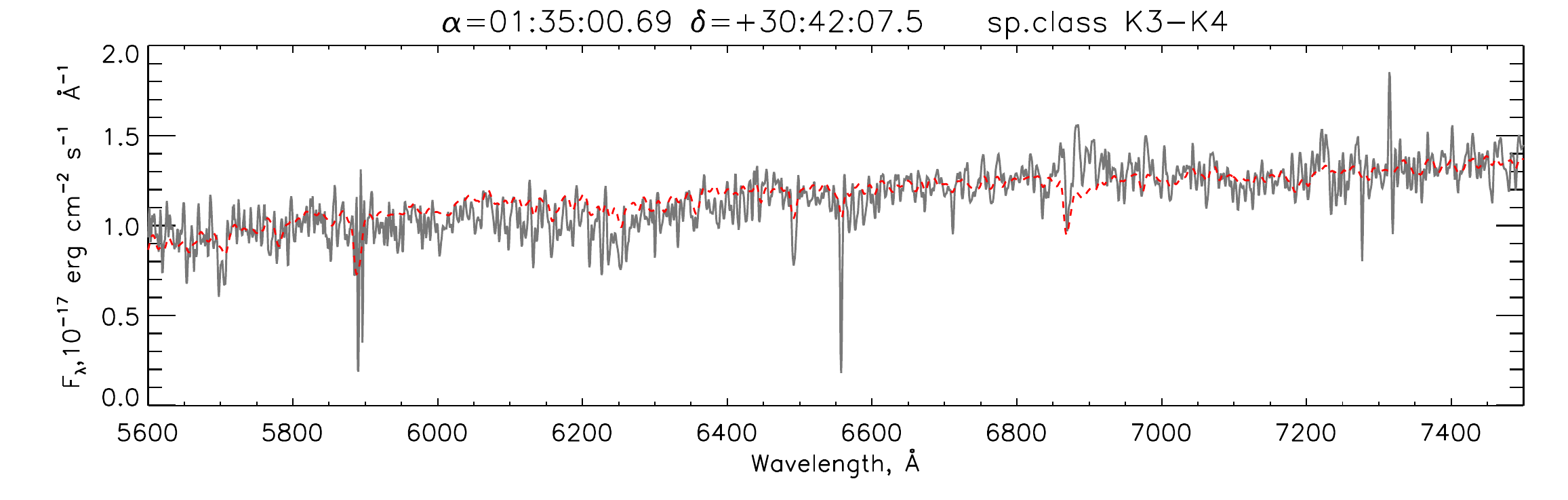}\\
\includegraphics[width=1.0\linewidth,clip,trim=30 5 0 0]{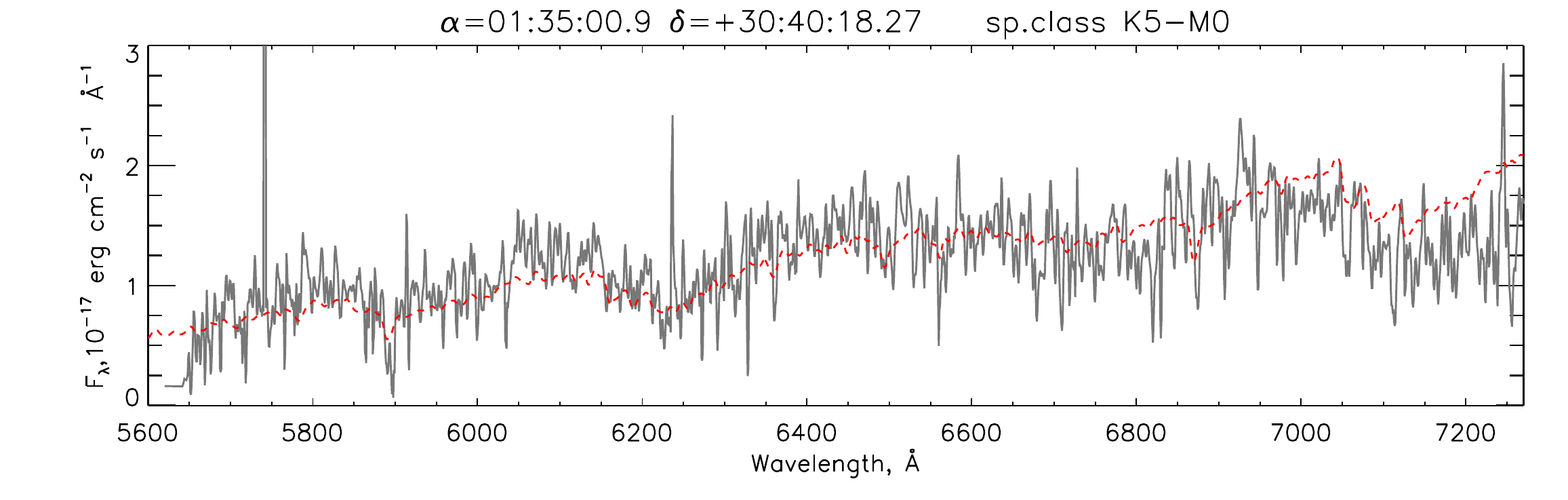}\\
\includegraphics[width=1.0\linewidth,clip,trim=30 5 0 0]{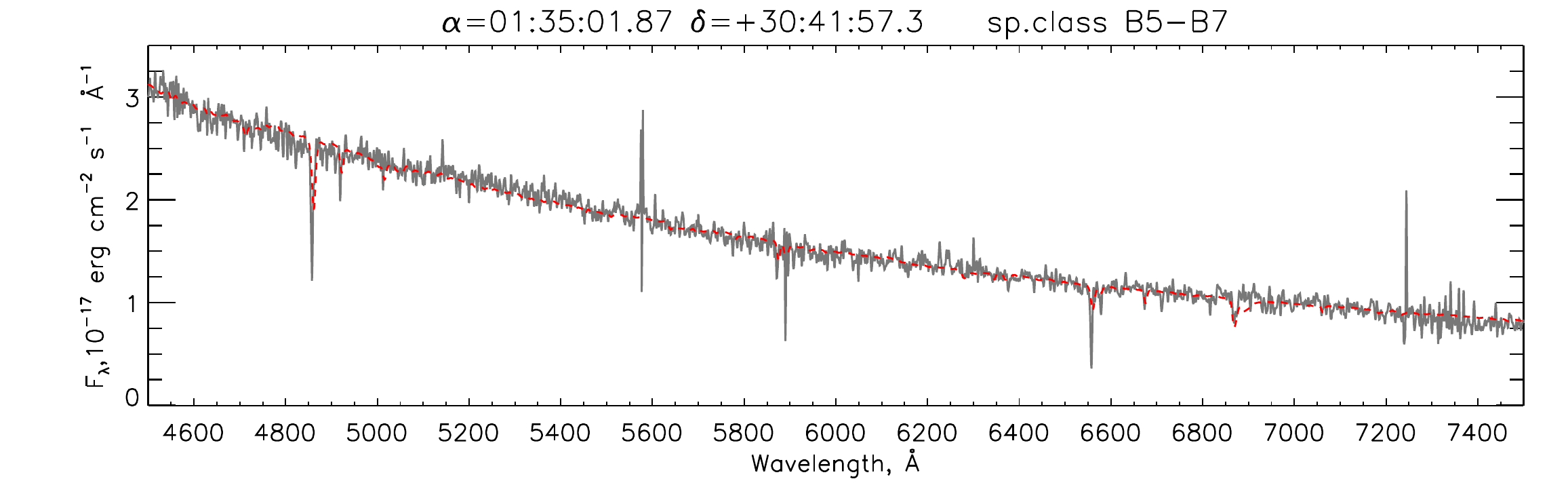}\\
\caption{From the top panel downwards, spectra of the stars: J013459.8+304156.98, J013500.69+304207.5, J013500.9+304018.27 and J013501.87+304157.3. For~comparison, the reddened spectra of HD\,42543 (M1\,Ia-ab), HD\,154733 (K3\,III), HD\,146051 (M0.5\,III) and HD\,164353 (B5\,Ib) are shown by red dashed lines.}
\label{star1}
\end{figure}
\unskip
\begin{figure}[H]
\includegraphics[width=1.0\linewidth,clip,trim=25 5 0 0]{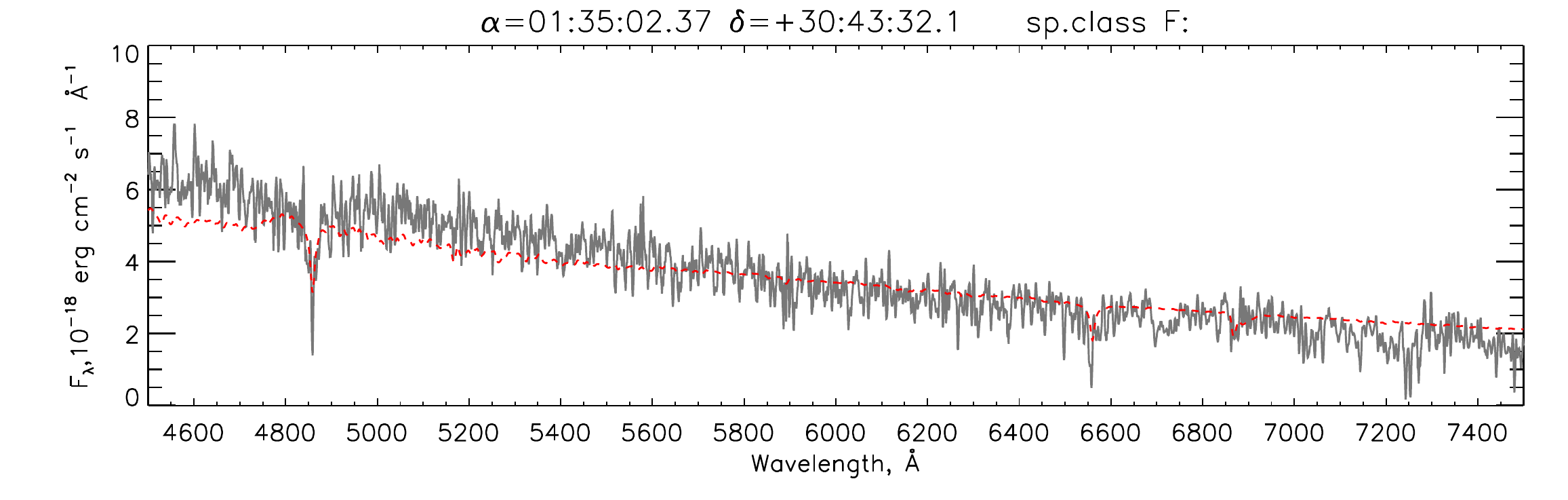}\\
\includegraphics[width=1.0\linewidth,clip,trim=25 5 0 0]{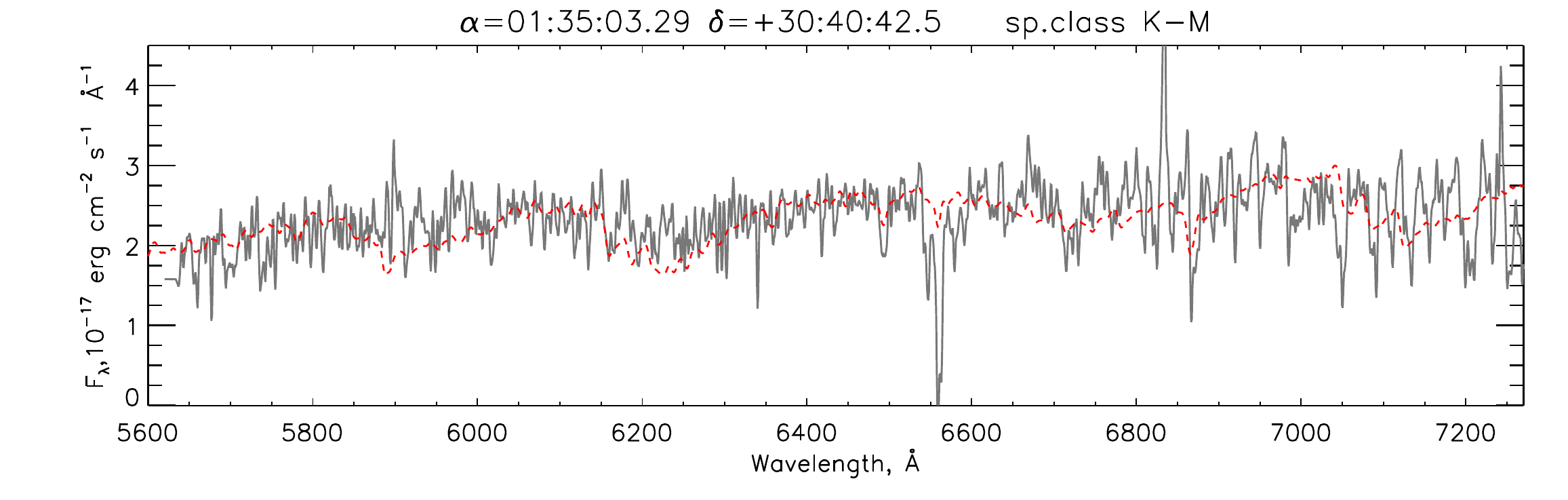}\\
\includegraphics[width=1.0\linewidth,clip,trim=25 5 0 0]{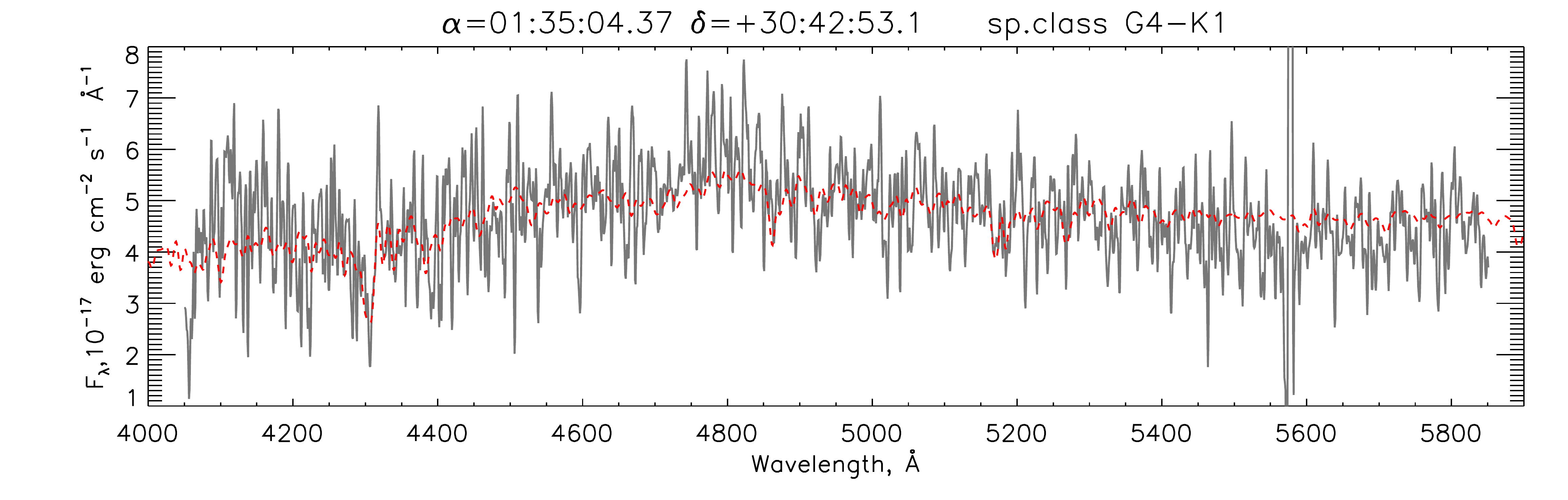}\\
\includegraphics[width=1.0\linewidth,clip,trim=25 5 0 0]{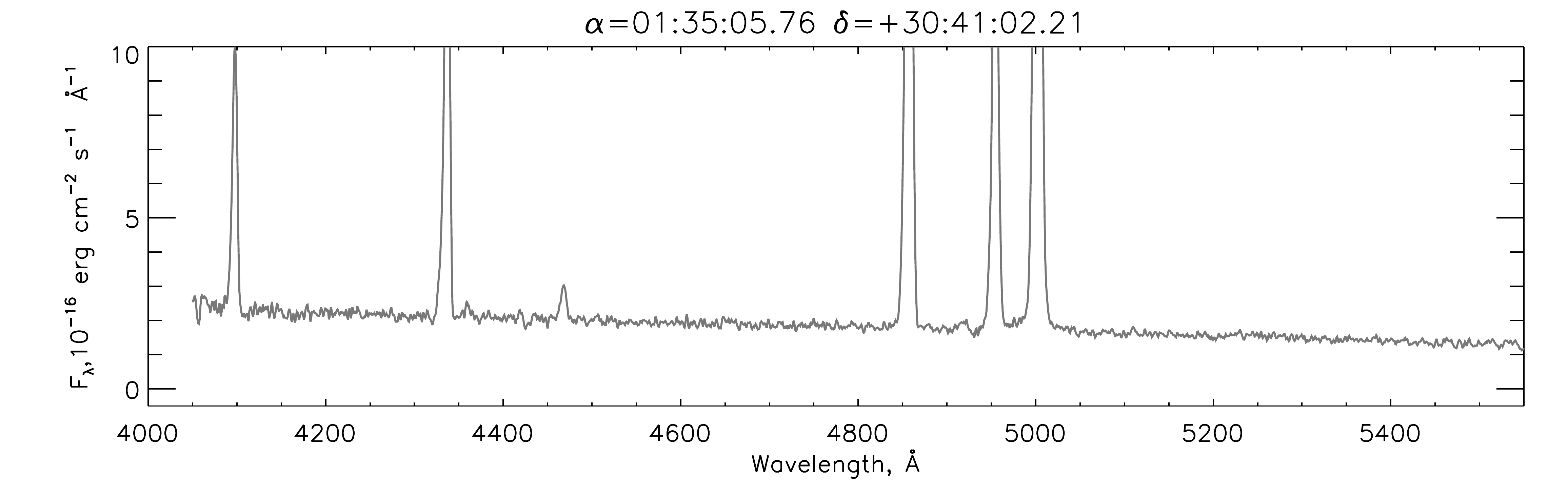}\\
\caption{From the top panel downwards, spectra of the stars:  J013502.37+304332.1, J013503.29+304042.5,  J013504.37+304253.1 and J013505.76+304102.21. For~comparison, the reddened spectra of HD\,128167 (F2\,V), HD\,102212 (M1\,III) and HD\,135722 (G8\,III)  are shown by red dashed~lines.  }
\label{star2}
\end{figure}
\unskip
\begin{figure}[H]
\includegraphics[width=1.0\linewidth,clip,trim=30 5 0 0]{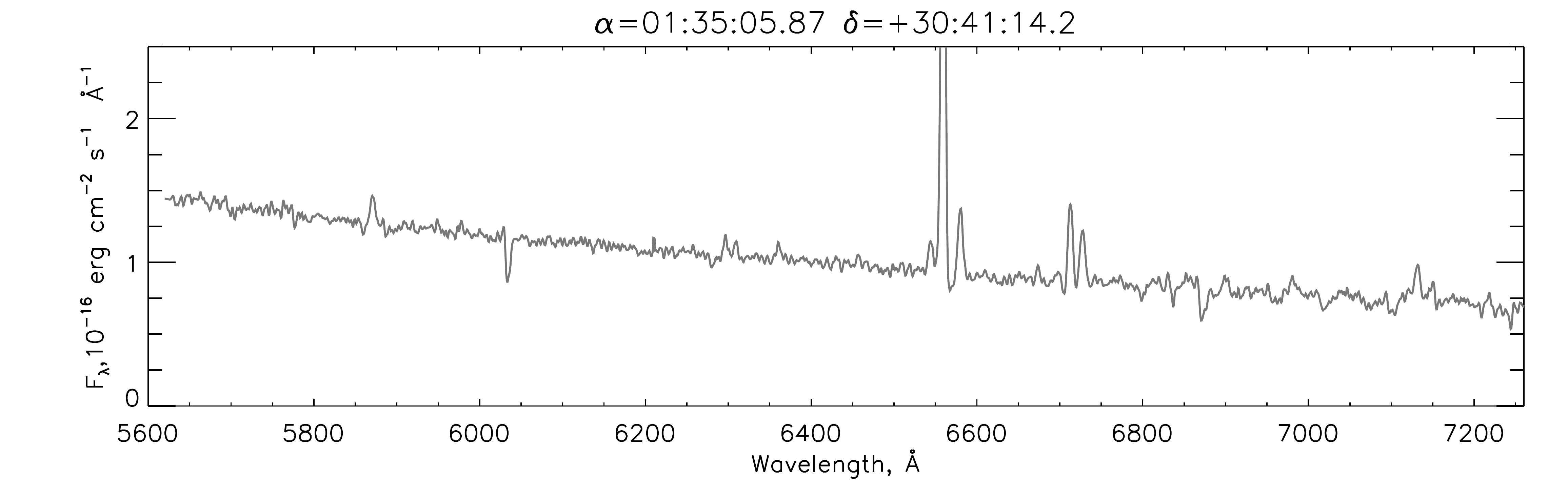}\\
\includegraphics[width=1.0\linewidth,clip,trim=30 5 0 0]{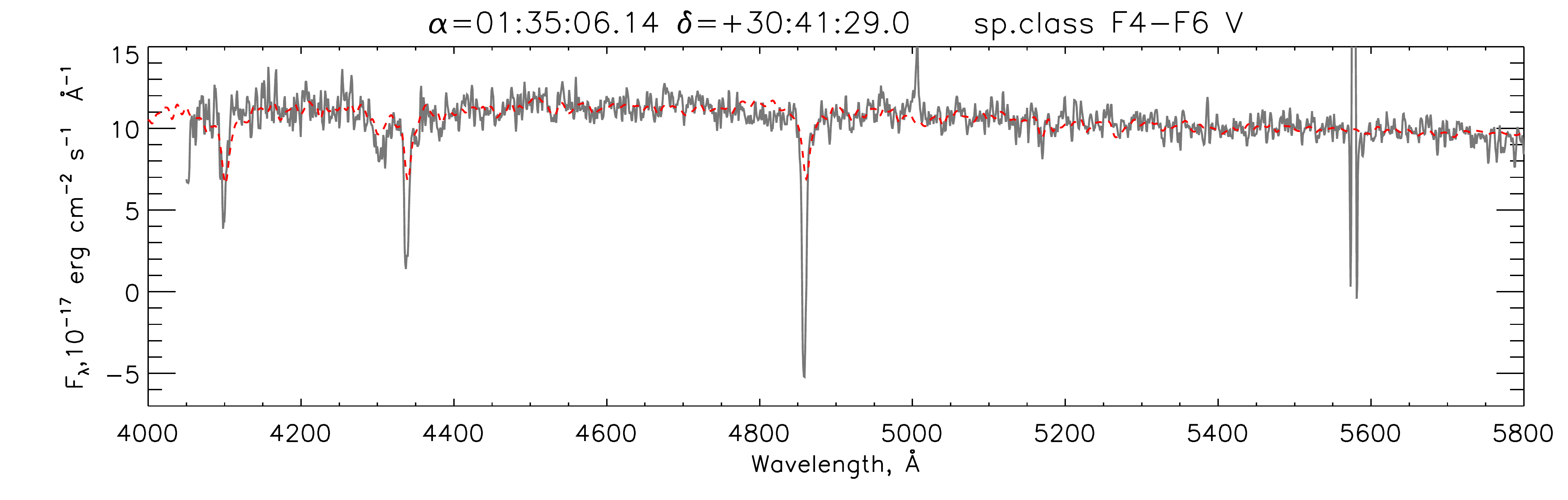}\\
\includegraphics[width=1.0\linewidth,clip,trim=30 5 0 0]{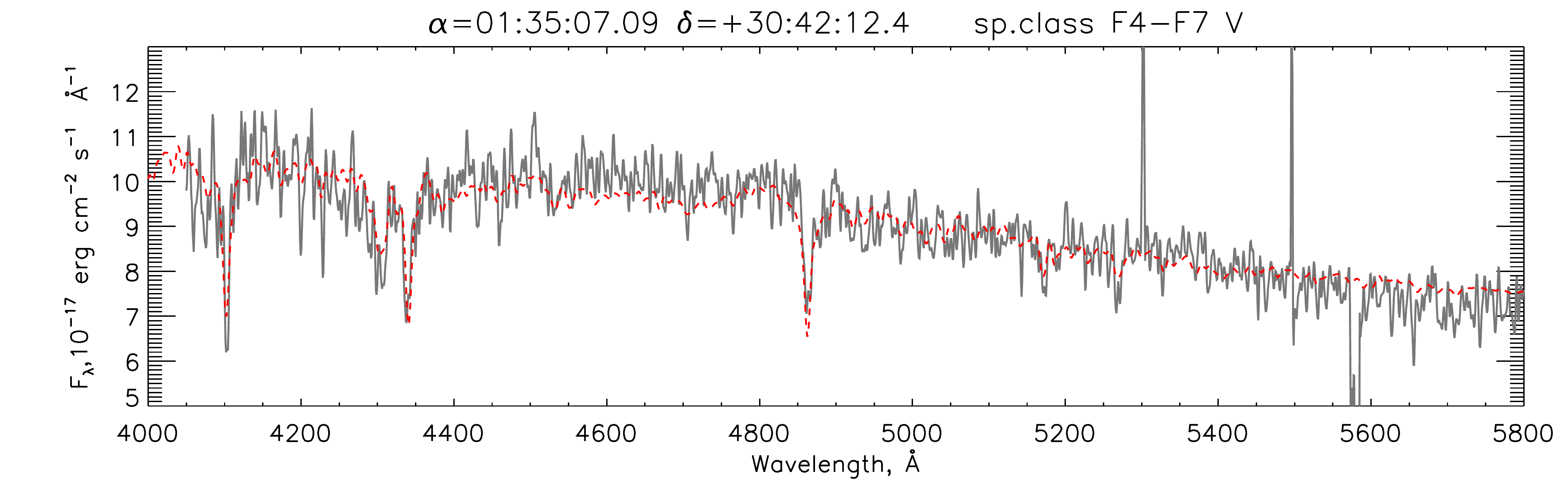}\\
\includegraphics[width=1.0\linewidth,clip,trim=30 5 0 0]{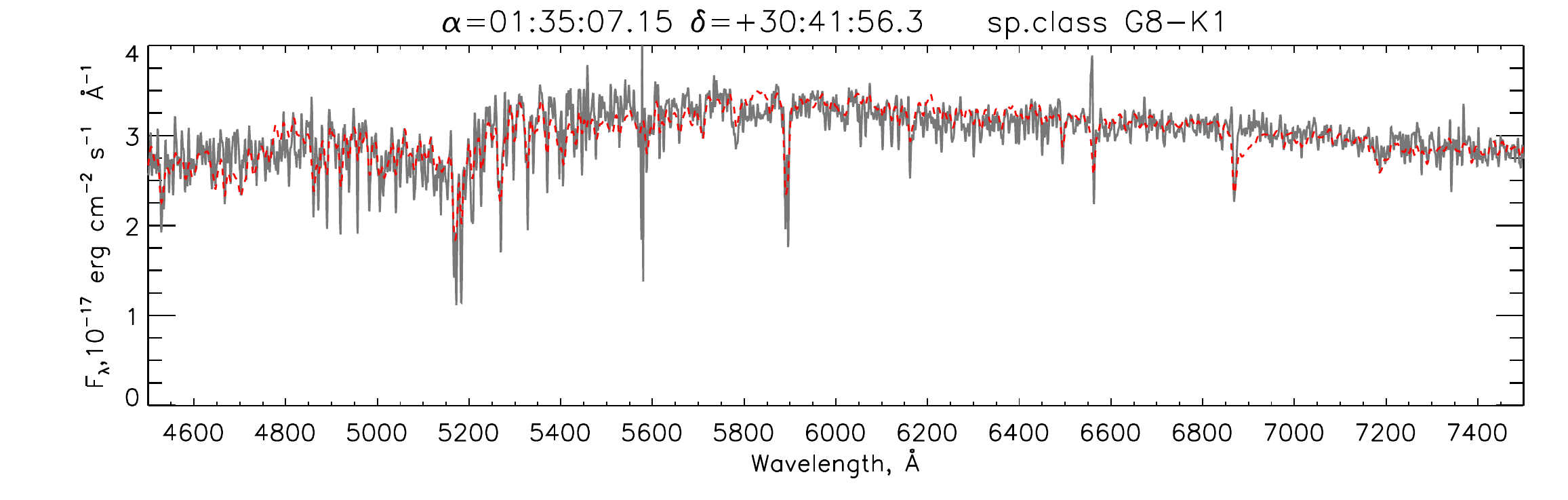}\\
\caption{From the top panel downwards, spectra of the stars: J013505.87+304114.2, J013506.14+304129.0, J013507.09+304212.4 and J013507.15+304156.3. For~comparison, the reddened spectra of HD\,126141 (F5\,V), HD\,101606 (F4\,V) and HD\,75532 (G8\,V) are shown by red dashed~lines.  }
\label{star3}
\end{figure}
\unskip
\begin{figure}[H]
\includegraphics[width=1.0\linewidth,clip,trim=30 5 0 0]{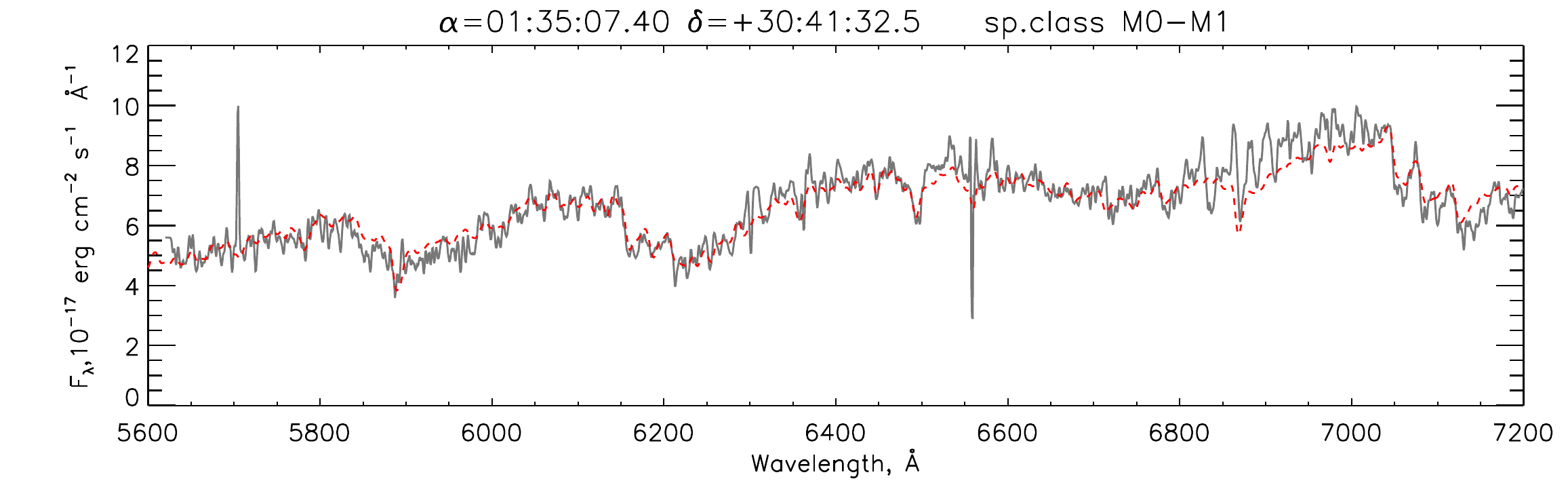}\\
\includegraphics[width=1.0\linewidth,clip,trim=30 5 0 0]{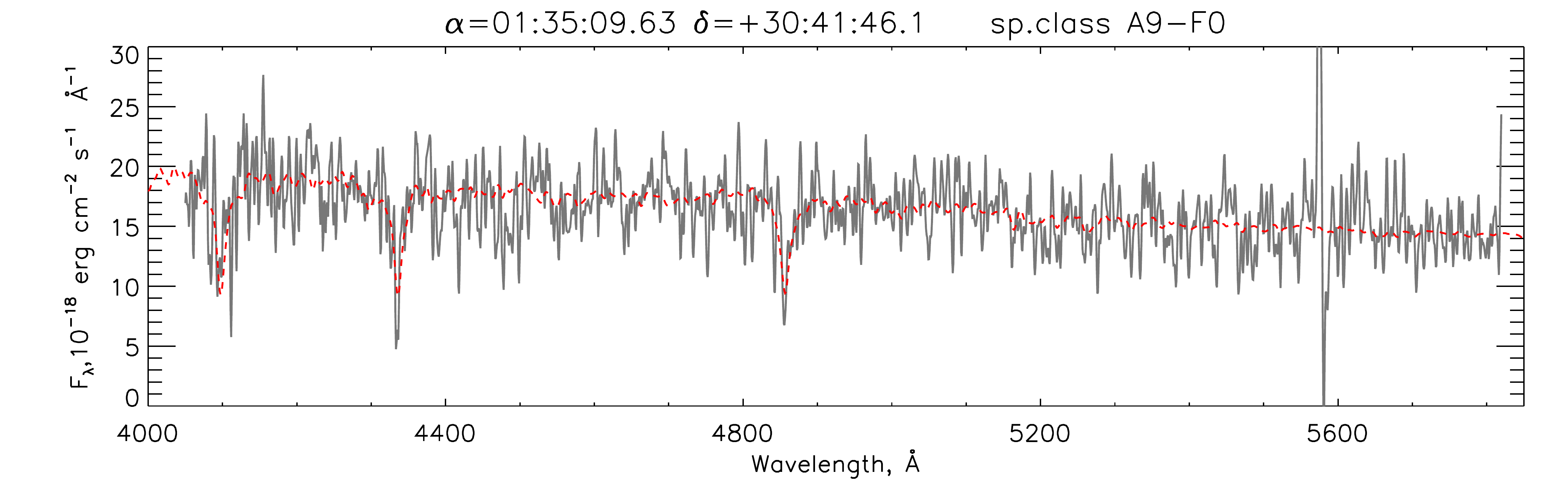}\\
\includegraphics[width=1.0\linewidth,clip,trim=30 5 0 0]{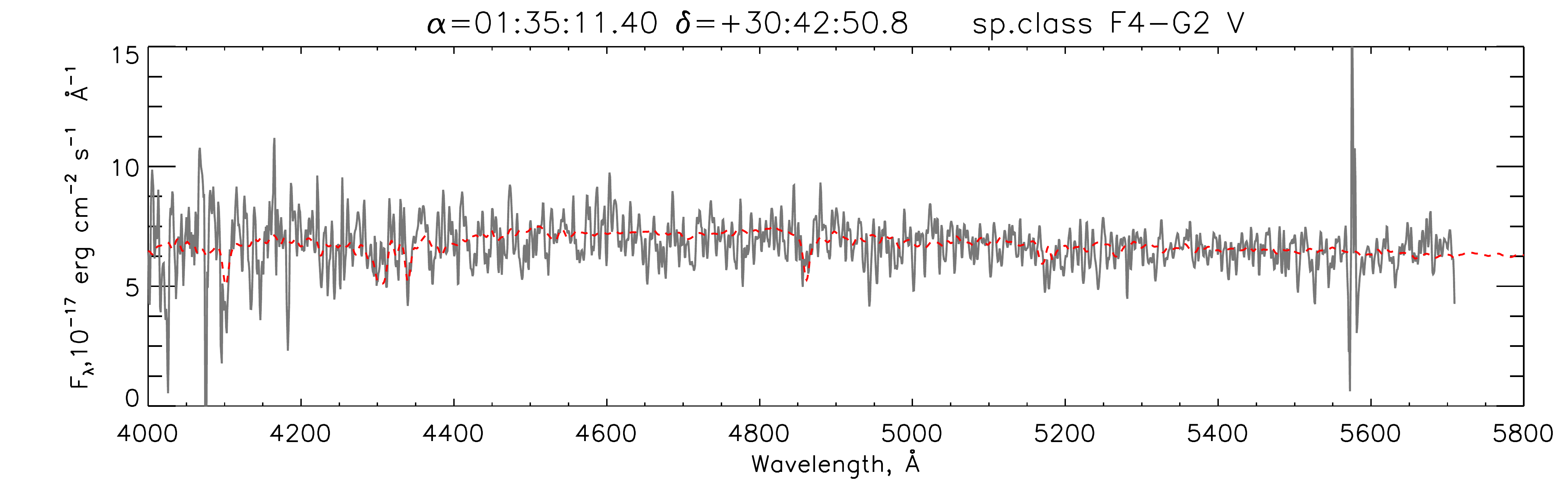}\\
\includegraphics[width=1.0\linewidth,clip,trim=30 5 0 0]{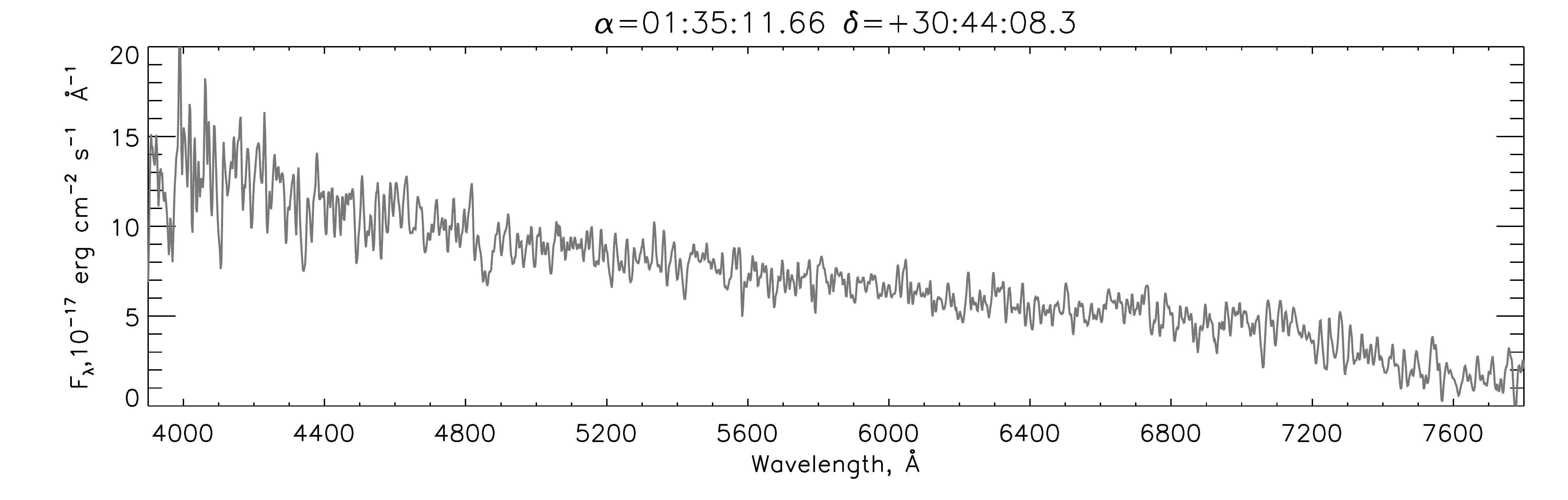}\\
\caption{From the top panel downwards, spectra of the stars: J013507.40+304132.5, J013509.63+304146.1, J013511.40+304250.8 and J013511.66+304408.3. For~comparison, the reddened spectra of HD\,146051 (M0.5\,III), HD\,50420 (A9\,III) and HD\,134169 (G1\,V) are shown by red dashed~lines.}
\label{star4}
\end{figure}
\unskip
\begin{figure}[H]
\includegraphics[width=1.0\linewidth,clip,trim=30 5 0 0]{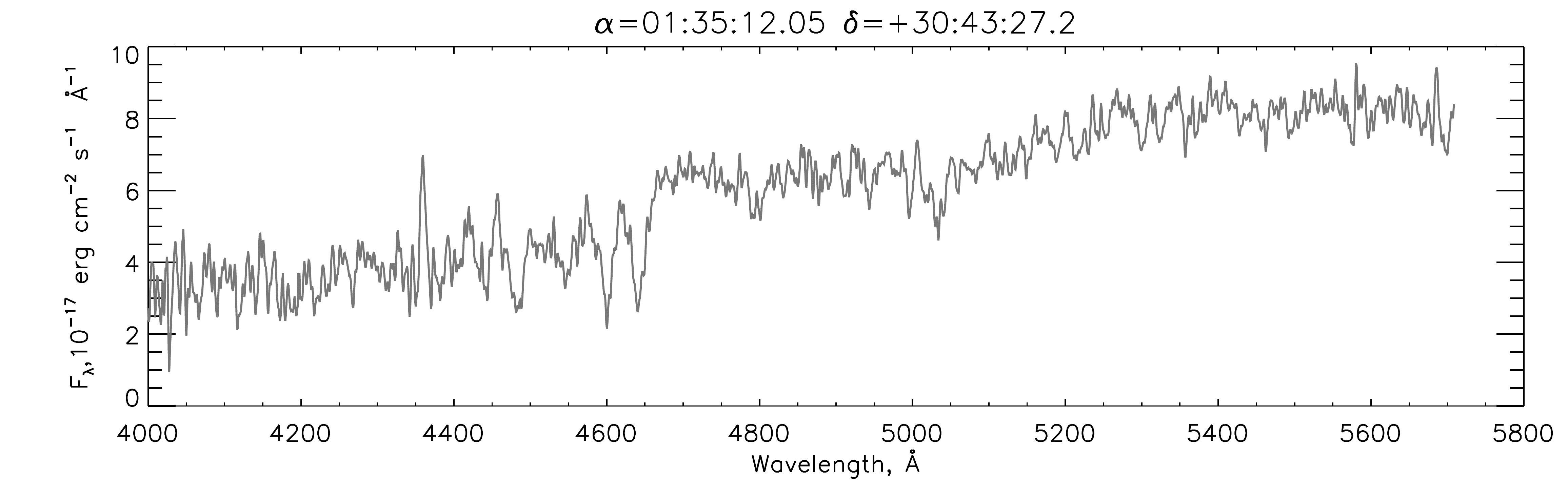}\\
\includegraphics[width=1.0\linewidth,clip,trim=30 5 0 0]{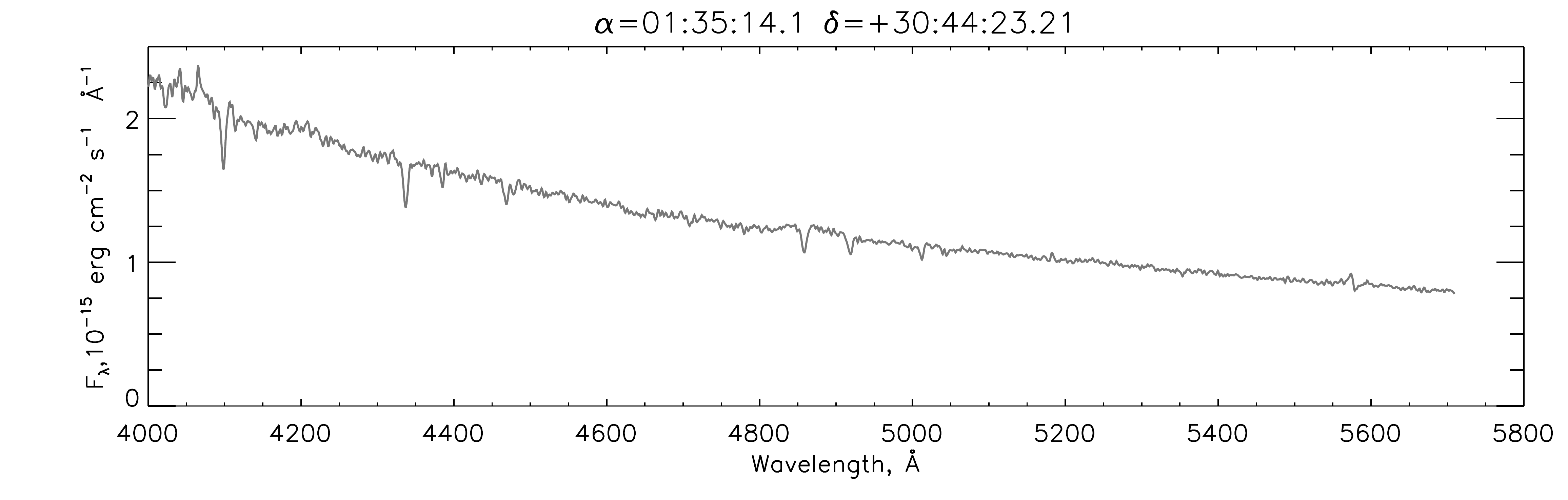}\\
\caption{From the top panel downwards, spectra of the stars: J013512.05+304327.2 and J013514.1+304423.21. }
\label{star5}
\end{figure}

\reftitle{References}



\end{document}